\documentclass[11pt,epsf,letterpaper]{article}
\usepackage{setspace}
\usepackage{color}
\usepackage{amsmath}
\usepackage{amsfonts}
\usepackage{verbatim}
\usepackage{amssymb}
\usepackage{graphicx,bm}
\usepackage{graphicx}
\usepackage{graphicx}
\usepackage{epstopdf}
\usepackage[caption=false]{subfig}

\definecolor{darkgreen}{rgb}{0,0.35,0}

\providecommand{\U}[1]{\protect\rule{.1in}{.1in}}
\begin{document}

\title{Analytic (3+1)-dimensional gauged Skyrmions, Heun and Whittaker-Hill
equations and resurgence}
\author{Fabrizio Canfora$^{1}$, Marcela Lagos$^{2}$, Seung Hun Oh$^{3}$,
Julio Oliva$^{2}$, Aldo Vera$^{2}$ \\
%EndAName
$^{1}$\textit{Centro de Estudios Cient\'{\i}ficos (CECS), Casilla 1469,
Valdivia, Chile}\\
$^{2}$\textit{Departamento de F\'{\i}sica, Universidad de Concepci\'{o}n,
Casilla 160-C, Concepci\'{o}n, Chile.}\\
$^{3}$\textit{Institute of Convergence Fundamental Studies, School of
Liberal Arts,} \\
\textit{Seoul National University of Science and Technology, Seoul 01811,
Korea}\\
{\small canfora@cecs.cl, marcelagos@udec.cl, shoh.physics@gmail.com,
juoliva@udec.cl, aldovera@udec.cl}}
\maketitle

\begin{abstract}
We show that one can reduce the coupled system of seven field equations of
the (3+1)-dimensional gauged Skyrme model to the Heun equation (which, for
suitable choices of the parameters, can be further reduced to the
Whittaker-Hill equation) in two non-trivial topological sectors. Hence, one
can get a complete analytic description of gauged solitons in (3+1)
dimensions living within a finite volume in terms of classic results in the
theory of differential equations and Kummer's confluent functions. We
consider here two types of gauged solitons: gauged Skyrmions and gauged
time-crystals (namely, gauged solitons periodic in time,\ whose time-period
is protected by a winding number). The dependence of the energy of the
gauged Skyrmions on the Baryon charge can be determined explicitly. The
theory of Kummer's confluent functions leads to a quantization condition for
the period of the time-crystals. Likewise, the theory of Sturm-Liouville
operators gives rise to a quantization condition for the volume occupied by
the gauged Skyrmions. The present analysis also discloses that resurgent
techniques are very well suited to deal with the gauged Skyrme model as
well. In particular, we discuss a very nice relation between the
electromagnetic perturbations of the gauged Skyrmions and the Mathieu
equation which allows to use many of the modern resurgence techniques to
determine the behavior of the spectrum of these perturbations.
\end{abstract}

\tableofcontents

\newpage

\section{Introduction}

In \cite{witten0} it was shown that the low energy limit of QCD can be
described by Skyrme theory \cite{skyrme}. The dynamical field of the Skyrme
action is an $SU(N)$ valued scalar field (here we will consider the $SU(2)$
case) whose topological solitons (called Skyrmions) describe Baryons. In
this context, the Baryon charge has to be identified with a suitable
topological invariant (see \cite{witten0}, \cite{finkrub}, \cite{manton}, 
\cite{skyrev0}, \cite{skyrev1}, \cite{giulini}, \cite{bala0}, \cite{ANW}, 
\cite{guada} and references therein).

Hence, not surprisingly, the Skyrme model is very far from being integrable
and, until very recently, no analytic solution with non-trivial topological
properties was known. One of the problematic consequences of this fact is
that the analysis of the phase diagram is quite difficult. In particular,
analytic results on finite density effects and on the role of the Isospin
chemical potential were unavailable despite the huge efforts in the
pioneering references \cite{klebanov}, \cite{chemical1}, \cite{chemical2}, 
\cite{chemical3}, \cite{chemical4}.

Even less is known on the (3+1)-dimensional gauged Skyrme model which
describes the coupling of a $U(1)$ gauge field with the Skyrme theory. The
importance, in many phenomenologically relevant situations\footnote{%
For the above reasons, when the coupling of Baryons with strong
electromagnetic fields cannot be neglected, the gauged Skyrme model comes
into play: its role is fundamental in nuclear and particle physics, as well
as in astrophysics.}, to analyze the interactions between Baryons, mesons
and photons makes mandatory the task to arrive at a deeper understanding of
the gauged Skyrme model (classic references are \cite{witten0}, \cite{Witten}
\cite{gipson}, \cite{goldstone}, \cite{dhoker}, \cite{rubakov}).

Obviously, the $U(1)$ gauged Skyrme model is ``even less integrable" than
the original Skyrme model. Consequently, as it was commonly assumed that to
construct analytic gauged Skyrmions was a completely hopeless goal, mainly
numerical tools were employed. Detailed numerical analysis of the gauged
Skyrme model in non-trivial topological sectors can be found in \cite%
{gaugesky1}, \cite{gaugesky2}\ and references therein.

Here it is worth to point out that to construct explicit topologically
non-trivial solutions is not just of academic interest. First, such
solutions allow to compute explicitly quantities of physical interest (such
as the mass spectrum and the critical Isospin chemical potential). Secondly,
once such solutions are available, one can test some modern ideas on how
non-perturbative configurations can improve usual perturbation theory. We
will come back on this issue in a moment.

Recently, in (\cite{canfora2}, \cite{canfora3}, \cite{canfora4}, \cite%
{canfora4.5}, \cite{yang1}, \cite{canfora6}, \cite{cantalla4}, \cite%
{cantalla5}, \cite{Giacomini:2017xno} and references therein) new
theoretical tools have been developed both in Skyrme and Yang-Mills theories
(see \cite{canYM1}, \cite{canYM2}, \cite{canYM3}\ and references therein),
which allow to build topologically non-trivial configurations even without
spherical symmetry.

The first (3+1)-dimensional analytic and topologically non-trivial solutions
of the Skyrme-Einstein system have been found in \cite{canfora6} using such
tools. Skyrmions living within a finite box in flat space-times have been
constructed using similar ideas in \cite{Fab1}: these results lead to the
derivation of the critical isospin chemical potential beyond which the
Skyrmion living in the box ceases to exist. Moreover, in the same reference,
it has been shown for the first time that the Skyrme model admits
Skyrmion-antiSkyrmion bound states\footnote{%
This is a very important result since, in particles physics, it is known
that Baryon-antiBaryon bound states do exist. From the Skyrme theory point
of view it is then necessary to prove that Skyrmion-antiSkyrmion bound
states do exist as well. This result has been achieved in \cite{Fab1}.}. In 
\cite{gaugsk}, using the results in \cite{canfora6}\ and \cite{Fab1}, a very
efficient method to build analytic and topologically non-trivial solutions
of the $U(1) $ gauged Skyrme model has been proposed: such a method will be
exploited here.

There are two types of topologically non-trivial gauged solitons which can
be obtained with the approach. The first type can be described as gauged
Skyrmions living within a finite volume. The second type are smooth
solutions of the $U(1)$ gauged Skyrme model whose periodic time-dependence
is topologically protected (this type of solitons can be called
topologically protected time-crystals).

Here we will show that both types of gauged solitons possess very special
configurations which allow a complete analytic description (which, in the
generic case, is not available) in terms of classic results in the theory of
differential equations. In these two special cases it is possible to reduce
the full system of seven coupled non-linear field equations of the gauged
Skyrme model in non-trivial topological sectors to the Heun equation (which,
in some cases, can be further reduced to the Whittaker-Hill equation). The
only difference between the Heun equations appearing in the gauged Skyrmions
and the gauged time-crystal sectors appears in the corresponding parameters.
This remarkable mapping allows to argue, among other things, that the time
period of the time-crystal is quantized (as it will be discussed in the
following sections). Likewise, Sturm-Liouville theory gives rise to a
quantization condition for the volume occupied by the gauged Skyrmions.
Moreover, we will also show that a particular type of electromagnetic
perturbations of these gauged solitons are described by the Mathieu equation
(another very well analyzed equation in mathematical physics).

This brings another surprising outcome of the present construction. Namely 
\textit{resurgence} does manifest itself in the (3+1)-dimensional gauged
Skyrme model as well. Resurgence (very nice physically-oriented reviews are 
\cite{res0}, \cite{res1}, \cite{res2} and \cite{res3}) is currently the main
framework which allows (at least in certain situations) to give a precise
mathematical sense (using suitable non-perturbative informations) to the
usual divergent perturbative expansions in quantum mechanics (QM) and
quantum field theory (QFT).

As it is well known, most of the perturbative expansions appearing in
theoretical physics are quite generically divergent. Resurgence helps in the
following way. The proliferation of Feynman diagrams leads to a factorial
growth of the perturbative coefficients and to the fact that the
perturbative series is, at most, an asymptotic series. One can try Borel
summation as a tool to give a meaning to such divergent series. However, in
all the physically interesting situations analyzed so far, the initial
divergent series becomes (through the Borel transform) a finite but
ambiguous expression (these ambiguities manifest themselves along suitable
lines in the complex $g$-plane, $g$ being the perturbative parameter).
Obviously, this situation is unsatisfactory as well. However, when one
analyzes theories with non-perturbative sectors one has to consider the
perturbative expansions in each of these non-perturbative sectors too (as
well as the fluctuations around them). It turns out that these
non-perturbative contributions are also ambiguous in the Borel sense.
Remarkably, as shown for the first time in the physical literature in \cite%
{res4} and \cite{res5} (as well as discussed in full generality in \cite%
{res6}), the non-perturbative ambiguities, at least in the models analyzed
in that references, exactly cancel those of the perturbative sector. Hence,
the perturbative divergence can be compensated by the non-perturbative
sectors. This is, roughly speaking, the resurgent paradigm. Starting from
the beautiful applications of resurgence techniques in \cite{res3}, there
has been a renewed interest on the applications of these ideas in
theoretical physics. Most of the results have been obtained in integrable
models, topological strings, QFT in 1+1 and 2+1 dimensions (in model such as 
\textbf{N=2} SUSY Yang-Mills theory and the Principal Chiral models; see 
\cite{res6.1}, \cite{Demulder:2016mja} and references therein) and in many
quantum mechanical problems (see \cite{dunne1}, \cite{Dunne:2016qix} and
references therein). Two classic ordinary differential equations in which
the resurgence paradigm works perfectly are the Mathieu and the
Whittaker-Hill equations\footnote{%
Besides the intrinsic interest of these two potentials, they often appear in
the reduction of quantum field theories in 1+1 dimensions on $%
%TCIMACRO{\U{211d} }%
%BeginExpansion
\mathbb{R}
%EndExpansion
\times S^{1}$\ (see \cite{Demulder:2016mja} and references therein) as well
as in the analysis of the Nekrasov-Shatashvili limit for the low-energy
behavior of N=2 supersymmetric SU(2) gauge theories (see \cite{dunne1} and
references therein).}.

On the other hand, there have been very few explicit expressions of
resurgence in non-integrable (3+1)-dimensional models. Thus, one may wonder
whether the appearance of resurgent behavior is, in a sense, generic or it
should be only expected in theories with a high degree of symmetries. The
present results provide with strong evidence supporting the first
hypothesis. Indeed, resurgence tools are very useful when analyzing the
spectrum of electromagnetic perturbations of these gauged solitons.

This paper is organized as follows: In the second section, a short review of
the gauged Skyrme model is presented. In the third section, gauged Skyrmions
and time-crystals are introduced. In the fourth section, the relations
between the Heun equation and the gauged solitons are explored. In Section %
\ref{conclusions}, we draw some concluding ideas.

\section{The $U(1)$ Gauged Skyrme Model}

\label{model}

The action of the $U(1)$ gauged Skyrme model in four dimensions is 
\begin{align}
S& =\int d^{4}x\sqrt{-g}\left[ \frac{K}{2}\left( \frac{1}{2}\mathrm{Tr}%
\left( R^{\mu }R_{\mu }\right) +\frac{\lambda }{16}\mathrm{Tr}\left( G_{\mu
\nu }G^{\mu \nu }\right) \right) -\frac{1}{4}F_{\mu \nu }F^{\mu \nu }\right]
\ ,  \label{sky1} \\
R_{\mu }& =U^{-1}D_{\mu }U\ ,\ \ G_{\mu \nu }=\left[ R_{\mu },R_{\nu }\right]
\ ,\ D_{\mu }=\nabla _{\mu }+A_{\mu }\left[ t_{3},\ .\ \right] \ ,
\label{sky2} \\
U& \in SU(2)\ ,\ \ R_{\mu }=R_{\mu }^{j}t_{j}\ ,\ \ t_{j}=i\sigma _{j}\ ,\
F_{\mu \nu }=\partial _{\mu }A_{\nu }-\partial _{\nu }A_{\mu }\ ,
\label{sky2.5}
\end{align}%
where $g$ is the the metric determinant, $A_{\mu }$ is the gauge potential, $%
\nabla _{\mu }$ is the partial derivative, the positive parameters $K$ and $%
\lambda $ are fixed experimentally and $\sigma _{j}$ are the Pauli matrices.
In our conventions $c=\hbar =\mu _{0}=1$, the space-time signature is $%
(-,+,+,+)$ and Greek indices run over space-time. The stress-energy tensor
is 
\begin{equation}
T_{\mu \nu }=-\frac{K}{2}\mathrm{Tr}\left[ R_{\mu }R_{\nu }-\frac{1}{2}%
g_{\mu \nu }R^{\alpha }R_{\alpha }\right. \,+\left. \frac{\lambda }{4}\left(
g^{\alpha \beta }G_{\mu \alpha }G_{\nu \beta }-\frac{g_{\mu \nu }}{4}%
G_{\sigma \rho }G^{\sigma \rho }\right) \right] +\bar{T}_{\mu \nu } \ , 
\notag  \label{timunu1}
\end{equation}%
with 
\begin{equation}
\bar{T}_{\mu \nu }=F_{\mu \alpha }F_{\nu }^{\;\alpha }-\frac{1}{4}F_{\alpha
\beta }F^{\alpha \beta }g_{\mu \nu } \ ,
\end{equation}%
being the electromagnetic energy-momentum tensor. The field equations read 
\begin{equation}
D^{\mu }\left( R_{\mu }+\frac{\lambda }{4}\left[ R^{\nu },G_{\mu \nu }\right]
\right) =0\ ,  \label{nonlinearsigma1}
\end{equation}%
\begin{equation}
\nabla _{\mu }F^{\mu \nu }=J^{\nu }\ ,  \label{maxwellskyrme1}
\end{equation}%
where $J^{\mu }$ is given by 
\begin{equation}
J^{\mu }=\frac{K}{2}\text{Tr}\left[ \widehat{O}R^{\mu }+\frac{\lambda }{4}%
\widehat{O}\left[ R_{\nu },G^{\mu \nu }\right] \right] \ ,  \label{current}
\end{equation}%
with%
\begin{equation*}
\widehat{O}=U^{-1}t_{3}U-t_{3}\ .
\end{equation*}%
It is worth to note that when the gauge potential reduces to a constant
along the time-like direction, the field equations (\ref{nonlinearsigma1})
describe the Skyrme model at a finite isospin chemical potential.

The term \textit{gauged Skyrmions} and \textit{gauged time-crystals} will
refer to smooth regular solutions of the coupled system in Eqs. (\ref%
{nonlinearsigma1}) and (\ref{maxwellskyrme1}) possessing a non-vanishing
winding number (defined below in Eq. (\ref{new4.1})).

\subsection{Gauged topological charge}

The standard parametrization of the $SU(2)$-valued scalar $U(x^{\mu })$ 
\begin{equation}
U^{\pm 1}(x^{\mu })=Y^{0}(x^{\mu })\mathbb{\mathbf{I}}\pm Y^{i}(x^{\mu
})t_{i}\ ,\ \ \left( Y^{0}\right) ^{2}+Y^{i}Y_{i}=1\,,  \label{standnorm}
\end{equation}%
where $\mathbb{\mathbf{I}}$ is the $2\times 2$ identity matrix and 
\begin{align}
Y^{0}& =\cos C \ ,\quad Y^{i}=n^{i}\cdot \sin C\ ,  \label{pions1} \\
n^{1}& =\sin F\cos G\ ,\ \ n^{2}=\sin F\sin G\ ,\ \ n^{3}=\cos F\ ,
\label{pions2}
\end{align}%
will be useful in the following computations.

The expression for the topological charge for the gauged Skyrme model has
been constructed in \cite{Witten} (see also the pedagogical analysis in \cite%
{gaugesky1}):%
\begin{equation}
W=\frac{1}{24\pi ^{2}}\int_{\Sigma }\rho _{B} \ ,  \label{new4.1}
\end{equation}
where 
\begin{equation}
\rho _{B}=\epsilon ^{ijk}\text{Tr}\left\{ \left( U^{-1}\partial _{i}U\right)
\left( U^{-1}\partial _{j}U\right) \left( U^{-1}\partial _{k}U\right)
-\partial _{i} \left[ 3A_{j}t_{3}\left( U^{-1}\partial _{k}U+\left( \partial
_{k}U\right) U^{-1}\right) \right] \right\} \ .  \label{new4.1.1}
\end{equation}%
There is an extra contribution with respect to the usual topological charge
in the Skyrme model which is responsible for the so-called Callan-Witten
effect \cite{Witten}.

The only case which is usually considered in the literature corresponds to a
space-like $\Sigma $. In these situations $W$ is the Baryon charge of the
configuration.

However, from the mathematical point of view, one can integrate the
three-form $\rho _{B}$ on any three-dimensional hypersurface and, in any
case, the number $W$ one obtains from Eq. (\ref{new4.1}) will be a
topological invariant. In particular, in \cite{Fab1} and \cite{gaugsk} it
has been shown that very interesting configurations are obtained when $%
\Sigma $ is time-like (the light-like case is also worth to be further
investigated). The interest of this case arises from the following
considerations. First of all, when $W\neq 0$ (no matter which hypersurface
one chooses) one cannot deform continuously the corresponding ansatz into $U=%
\mathbb{\mathbf{I}}$. Therefore, when $\rho _{B}$ is different from zero
along a time-like hypersurface so that $\Sigma $ must be time-like in order
to get $W\neq 0$ one gets non-trivial gauged solitons which depend on time
(otherwise $\rho _{B}$ would vanish along time-like hypersurfaces).
Moreover, the time-dependence of these gauged solitons is topologically
protected since, by homotopy theory, $W$ cannot change under continuous
deformations\ and this implies that they cannot decay into static
configurations (since, for static configurations, $\rho _{B}$ vanishes along
a time-like hypersurface). Since it turns out that these gauged solitons are
periodic in time they can be called topologically protected time crystals.
To the best of authors knowledge, the examples constructed in the following
sections are the first time-periodic solutions whose time-period is
topologically protected by homotopy theory. As it will be discussed in the
following sections, classic results in the theory of Kummer's confluent
functions determine the allowed time-periods.

\section{Gauged Skyrmions and Time Crystals}

Here we will describe the theoretical tools introduced in \cite{gaugsk},
which are needed to build the novel gauged Skyrmions and the gauged
time-crystals which will be analyzed in the following sections. Since one of
the main physical motivations behind the analysis in \cite{Fab1}, \cite%
{gaugsk} was to study finite volume effects, the first step of the analysis
is to put the gauged Skyrme model within a box of finite volume. The easiest
way to achieve this goal is to introduce the following flat metric 
\begin{equation}
ds^{2}=-dt^{2}+l^{2}\left( dr^{2}+d\gamma ^{2}+d\phi ^{2}\right) \ .
\label{Minkowski}
\end{equation}%
The length scale $l$ represents the size of the box. The dimensionless
coordinates $r$, $\gamma $ and $\phi $ have the periods: 
\begin{equation}
0\leq r\leq 2\pi \ ,\quad 0\leq \gamma \leq 4\pi \ ,\quad 0\leq \phi \leq
2\pi \ .  \label{period0}
\end{equation}

The following parametrization of the $SU(2)$-valued scalar $U$ will be
considered: 
\begin{equation}
U=e^{t_{3}\alpha }e^{t_{2}\beta }e^{t_{3}\rho } \ ,  \label{euan}
\end{equation}%
where $\alpha $, $\beta $ and $\rho $ are the Euler angles which in a single
covering of space take the values $\alpha \in \lbrack 0,2\pi ]$, $\beta \in
\lbrack 0,\frac{\pi }{2}]$ and $\rho \in \lbrack 0,\pi ]$.

\subsection{Gauged Skyrmions}

The ansatz for the gauged Skyrmion can be chosen as 
\begin{equation}
\alpha =p\frac{\gamma }{2} \ ,\quad \beta =H(r) \ , \quad \rho =q\frac{\phi 
}{2} \ , \quad p,\ q\in 
%TCIMACRO{\U{2115} }%
%BeginExpansion
\mathbb{N}
%EndExpansion
\ .  \label{ans1}
\end{equation}%
In order for the ansatz to cover $SU(2)$ an integer number of times, the two
parameters $p$ and $q\ $\ must be integer.

The profile $H$ must be static, and the electromagnetic potential has to be
chosen as 
\begin{equation}
A_{\mu }=(b_{1}(r),0,b_{2}(r),b_{3}(r))\ .  \label{EMpotans1}
\end{equation}

As it has been shown in \cite{gaugsk}, if one requires the following two
conditions 
\begin{equation}
X_{1}=-\frac{\lambda (p^{2}+q^{2})}{2}=\text{constant}\ ,  \label{condin1}
\end{equation}%
where 
\begin{equation}
X_{1}(r):=4\lambda \left(
-2l^{2}b_{1}^{2}+b_{2}(2b_{2}+p)+b_{3}(2b_{3}-q)\right) \ ,  \label{prex1}
\end{equation}%
and 
\begin{equation}
b_{2}(r)=-\frac{q}{p}b_{3}(r)-\frac{p^{2}-q^{2}}{4p}\ ,  \label{condin2}
\end{equation}%
then\textit{\ the coupled Skyrme Maxwell system} made by Eqs. (\ref%
{nonlinearsigma1}) and (\ref{maxwellskyrme1}) in a topologically non-trivial
sector \textit{can be reduced consistently to the following system of two
ODEs} below 
\begin{equation}
\left( \frac{8l^{2}}{p^{2}+q^{2}}+2\lambda \cos ^{2}(H)\right) H^{\prime
\prime }+\sin (2H)\left( l^{2}-\lambda H^{\prime 2}\right) =0\ ,
\label{fullsk1}
\end{equation}
\begin{equation}
b_{3}^{\prime \prime }-\frac{K}{4}(q-4b_{3})\sin ^{2}(H)\left(
4l^{2}+4\lambda H^{\prime 2}+\lambda \left( p^{2}+q^{2}\right) \cos
^{2}(H)\right) =0\ .  \label{fullm1}
\end{equation}

In other words, the three coupled gauged Skyrme equations in Eq. (\ref%
{nonlinearsigma1}) and the corresponding four Maxwell equations in Eq. (\ref%
{maxwellskyrme1}) with the Skyrme ansatz in Eqs. (\ref{euan}) and (\ref{ans1}%
) and the gauge potential in Eq. (\ref{EMpotans1}) reduce to Eqs. (\ref%
{fullsk1}) and (\ref{fullm1}) when the two algebraic conditions in Eqs. (\ref%
{condin1}) and (\ref{condin2}) are satisfied. This is a remarkable
simplification and below we will show that indeed this sector contains
physically relevant configurations.

Consequently, in order to construct explicitly gauged Skyrmions, the optimal
strategy is to determine the Skyrme profile $H(r)$ from Eq. (\ref{fullsk1})
and then, once $H(r)$ is known, Eq. (\ref{fullm1}) becomes a linear
Schrodinger-like equation for the component $b_{3}(r)$ of gauge potential.
Once $b_{3}$ is known, the other two components of the gauge potential are
determined by the two algebraic relations in Eqs. (\ref{condin1}) and (\ref%
{condin2}).

It is worth to note that, generically, once Eq. (\ref{fullsk1}) is solved in
terms of suitable elliptic integrals, the resulting Eq. (\ref{fullm1})
(despite being a linear equation) will not have explicit analytic solutions.
The reason is that the effective potential one gets replacing the solution
of Eq. (\ref{fullsk1}) into Eq. (\ref{fullm1}) will not be a solvable
potential in the generic case.

In fact, in the following sections we will analyze the most elegant
solutions of the above sector in which it is possible to obtain a complete
analytic construction of gauged Skyrmions in terms of classic results in the
theory of differential equations.

%\subsubsection{Topological charge}

The Baryon number corresponding to the Skyrme ansatz in Eqs. (\ref{euan})
and (\ref{ans1}) and to the gauge potential in Eq. (\ref{EMpotans1}) is 
\begin{equation}
B=\int B_{0}drd\gamma d\phi =-p\,q\int \sin (2H)dH+2\Big[\cos
^{2}(H(r))\left( q\,b_{2}(r)-p\,b_{3}(r)\right) \Big]_{0}^{2\pi },
\end{equation}%
and it leads to 
\begin{subequations}
\label{topchsk}
\begin{align}
B=& -pq-2\left( q\,b_{2}(0)-p\,b_{3}(0)\right) = -\frac{(p^2+q^2)(q-4b_3(0))%
}{2p} \ , \\
B=& pq+2\left( q\,b_{2}(2\pi )-p\,b_{3}(2\pi )\right) = \frac{%
(p^2+q^2)(q-4b_3(2\pi))}{2p} \ ,
\end{align}%
depending on the boundary values that we assume: $H(2\pi )=\pi /2$, $H(0)=0$
or $H(2\pi )=0$, $H(0)=\pi /2$ respectively.

It is worth to point out that the gauged Skyrmions which will be constructed
here are periodic in two spatial directions (namely $\gamma$ and $\phi$),
while satisfy Dirichlet boundary condition in the coordinate $r$ (as it is
clear from the fact that the values of $H(r)$ at $r=0$ and $r=\frac{\pi}{2}$
are fixed). Therefore, the spatial topology does not correspond to a $3$%
-torus but rather to a ($2$-torus)$\times$(finite interval), where the $2$%
-torus corresponds to the $\gamma$ and $\phi$ coordinates while the finite
interval corresponds to the $r$ coordinate.

\subsection{Gauged time crystal}

Quite recently Wilczek and Shapere \cite{timec1}, \cite{timec2}, \cite%
{timec3}, asked the following very intriguing question (both in classical
and in quantum physics): is it\textit{\ possible to spontaneously break time
translation symmetry in physically sensible models?}

These questions, until very recently, have been mainly analyzed in condensed
matter physics. It is well known that powerful no-go theorems \cite{timec4}, 
\cite{timec5} severely restrict the concrete realization of time-crystals.
Novel versions of the original ideas (see \cite{timec1}, \cite{timec2}, \cite%
{timec3}; a nice review is \cite{timecr}) allowed to realize time-crystals
in ``solid-states" settings (see \cite{timec5.5}, \cite{timec5.6}, \cite%
{timec5.7}, \cite{timec6}, \cite{timec7}, \cite{timec9} and references
therein).

On the other hand, it seems that the first examples of time-periodic
solutions which are topologically protected by homotopy theory in nuclear
and particles physics have been found in \cite{Fab1}, \cite{gaugsk}. The key
ingredients of the gauged Skyrme model is the possibility to have
configurations with non-vanishing winding number along time-like
hypersurfaces. Thanks to this fact, one can construct analytic time-periodic
configurations which cannot be deformed continuously to the trivial vacuum
as they possess a non-trivial winding number. Moreover, homotopy theory
ensures that such solitons can only be deformed into other solitons with the
same time period. Hence, these configurations can decay only into other
time-periodic configurations. For these reasons, the name \textit{%
topologically protected time crystals} is appropriate.

We will consider the line element in Eq. (\ref{Minkowski}). The Skyrme
ansatz in this case is 
\end{subequations}
\begin{equation}
\alpha =\frac{\phi }{2}\ ,\quad \beta =H(r)\ ,\quad \rho =\frac{\omega t}{2}%
\ ,  \label{eultc}
\end{equation}%
where $\omega $ is a frequency so that $\rho $ is dimensionless (as it
should be). One can see that, with the above ansatz, the topological density 
$\rho _{B}$ in Eq. (\ref{new4.1.1}) has a term proportional to%
\begin{equation*}
\rho _{B}\sim dt\wedge dr\wedge d\phi +... \ ,
\end{equation*}%
where the dots represent terms which depend on $A_{\mu }$.

As far as the ansatz for the gauged-time crystal is concerned, it basically
corresponds to a Wick rotation of the gauged Skyrmion in which one takes one
of the two spatial periodic coordinates (which, for the gauged Skyrmions,
are $\gamma$ and $\phi$) as time-like coordinate. Because of this, the
dependence of the time-crystal on time is necessarily periodic. This
explains why, in the case of the time-crystal, one can compute the winding
number along the three-dimensional time-like hypersurface corresponding to
the coordinates $t$, $\phi$ and $r$. The time-integration domain in the
three-dimensional integral defining the winding number in the gauged
time-crystal case corresponds to one period.

Thus, the above ansatz is a good candidate to be a time-crystal since its
topological density can be integrated along a time-like hypersurface. The
electromagnetic potential has the form \eqref{EMpotans1}, but the coordinate
ordering is 
\begin{equation}
x^{\mu }=(\gamma ,r,t,\phi )\ .  \label{cotc}
\end{equation}%
In this case as well, as it has been shown in \cite{gaugsk}, if the
following two relations among the three components of the gauge potential in
Eq. (\ref{EMpotans1}) 
\begin{equation}
X_{2}=\lambda \left( l^{2}\omega ^{2}-1\right) =\text{constant}\ ,
\label{condin3}
\end{equation}%
where 
\begin{equation}
X_{2}(r):=8\lambda \left( l^{2}b_{1}(\omega
-2b_{1})+2b_{2}^{2}+b_{3}(1+2b_{3})\right) \ ,
\end{equation}%
and 
\begin{equation}
b_{3}(r)=l^{2}\omega b_{1}(r)-\frac{l^{2}\omega ^{2}}{4}-\frac{1}{4}\ ,
\label{condin4}
\end{equation}%
hold, then \textit{\ the coupled Skyrme Maxwell system} made by Eqs. (\ref%
{nonlinearsigma1}) and (\ref{maxwellskyrme1}) in a topologically non-trivial
sector \textit{can be reduced consistently to the following system of two
ODEs} below: 
\begin{equation}
2\left( \lambda \left( l^{2}\omega ^{2}-1\right) \cos ^{2}(H)-4l^{2}\right)
H^{\prime \prime }+\left( l^{2}\omega ^{2}-1\right) \sin (2H)\left(
l^{2}-\lambda H^{\prime 2}\right) =0\ ,  \label{profredtc}
\end{equation}
\begin{equation}
b_{1}^{\prime \prime }+\frac{K}{8}(\omega -4b_{1})\sin ^{2}(H)\left(
l^{2}(\lambda \omega ^{2}-8)-\lambda +\lambda \left( l^{2}\omega
^{2}-1\right) \cos (2H)-8\lambda H^{\prime 2}\right) =0\ .  \label{b1redtc}
\end{equation}
The optimal strategy is then to determine the Skyrme profile $H(r)$ from Eq.
(\ref{profredtc}). Once $H(r)$ is known, Eq. (\ref{b1redtc}) becomes a
linear Schrodinger-like equation for the gauge potential component $b_{1}(r)$%
. The other components of the gauge potential are determined by solving the
simple algebraic conditions in Eqs. (\ref{condin3}) and (\ref{condin4}).

Generically, once Eq. (\ref{profredtc}) is solved in terms of suitable
elliptic integrals, the resulting Eq. (\ref{b1redtc}) will not have explicit
analytic solutions (despite being a linear equation). The reason is that the
effective potential one gets replacing the solution of Eq. (\ref{profredtc})
into Eq. (\ref{b1redtc}) will not be solvable.

In fact, in the following sections we will analyze the most elegant
time-crystals in which it is possible to obtain a complete analytic
construction in terms of classic results in the theory of differential
equations.

As we did in the previous section for the gauged Skyrmion, we also calculate
here for the time crystal the non vanishing winding number, that is

\begin{align}
W = \int B_{2}drd\left( \omega \gamma \right) d\phi = & 1+2\Big[\cos
^{2}(H(r))\left( \frac{b_{1}(r)}{\omega }-b_{3}(r)\right) \Big]_{0}^{2\pi} 
\notag \\
= & 1-2\left( \frac{b_{1}(0)}{\omega }-b_{3}(0)\right)  \notag \\
= & \frac{(1-l^2\omega^2)\left(\omega-4b_1(0)\right)}{2\omega} \ ,
\label{windingtc}
\end{align}%
if we consider $r\in \lbrack 0,2\pi ]$, $\omega \gamma \in \lbrack 0,4\pi ]$%
, $\phi \in \lbrack 0,2\pi ]$ and $H(2\pi )=\pi /2$, $H(0)=0$.

However, a normal topological charge is also present here due to the
correction from the electromagnetic potential. By taking $B_{0}$ as defined
in \eqref{new4.1.1} as an integral over spatial slices, we obtain 
%\begin{equation}
\begin{align*}
B=\int B_{0}drdzd\phi = -2\Big[\cos ^{2}(H(r))b_{2}(r)\Big]_{0}^{2\pi} =
2b_{2}(0) = \frac{l}{2}(\omega-4 b_1(0)) \sqrt{1-l^2\omega^2} \ ,
\end{align*}
with the same boundary values used as in \eqref{windingtc}, with the
difference now that we have $z$ in place of $\gamma $ for which we consider $%
z\in \lbrack 0,2\pi ]$. The charge $B$ is non zero as long as $b_{2}(0)\neq
0 $.

In the following sections we will analyze the most elegant solutions of the
above sector.

\subsection{Extended duality}

In this subsection it is shown that a sort of electromagnetic duality exists
between the gauged Skyrmion and the gauged time-crystal constructed above.
In order to achieve this goal, it is useful to observe that a mapping
between the time crystal to the Skyrmion should involve a transformation 
\begin{equation}
\gamma \rightarrow i\,l\,\gamma \ , \qquad z\rightarrow \frac{i}{l}z \ ,
\label{imtrmetr}
\end{equation}%
so that the signature can be changed appropriately. Then, it is an easy task
to see that Eq. (\ref{profredtc}) and Eq. (\ref{b1redtc}) are mapped to Eq. (%
\ref{fullsk1}) and Eq. (\ref{fullm1}) under the linear transformation 
\begin{equation}
a_{1}=\frac{i}{l}b_{2} \ , \quad a_{2}=i\,l\,b_{1} \ , \quad a_{3}=-b_{3} \ .
\label{dualtr}
\end{equation}%
The appearance of the imaginary units is not alarming, since one also needs
a coordinate transformation like \eqref{imtrmetr} to map the one space-time
metric to the other. Notice that the imaginary part of the transformation
involves only the $\gamma $ and $z$ components of $A_{\mu }$. Hence, the end
result after utilizing \eqref{imtrmetr} is a real electromagnetic tensor of
the Skyrmion case.

In the following two tables we gather the electromagnetic potentials of the
Skyrmion and of the time crystal (T.C.) as well the necessary
transformations that need to be made - not only to the field but also to
parameters and variables - in order to make the transition from the one case
to the other. 
\begin{equation*}
\begin{tabular}{|c|c|c|c|}
\hline
{\ } & $A_{\mu }$ & \text{coordinate system} $x^{\mu }$ & \text{time variable%
} \\ \hline
\text{Skyrmion} & $(b_{1}(r),0,b_{2}(r),b_{3}(r))$ & $(z,r,\gamma ,\phi )$ & 
$z$ \\ \hline
\text{T. C. before transformation} & $(a_{1}(r),0,a_{2}(r),a_{3}(r))$ & $%
(\gamma ,r,z,\phi )$ & $\gamma $ \\ \hline
\end{tabular}%
\end{equation*}%
\begin{equation*}
\begin{tabular}{|c|c|}
\hline
\multicolumn{2}{|l|}{\hspace{.3in}\text{T.C.} \hspace{.1in} $\longrightarrow 
$ \hspace{.07in} \text{Skyrmion}} \\ \hline
$a_{1}(r)$ \hspace{1pt} & $i\,b_{2}(r)/l$ \\ \hline
$a_{2}(r)$ \hspace{1pt} & $i\,l\,b_{1}(r)$ \\ \hline
$a_{3}(r)$ \hspace{1pt} & $-b_{3}(r)$ \\ \hline
$\omega $ \hspace{1pt} & $-i/l$ \\ \hline
$(\gamma ,z)$ \hspace{1pt} & $(i\,l\,\gamma ,iz/l)$ \\ \hline
$(E_{1},B_{2}, B_{3})$ \hspace{1pt} & $(-B_{3}, - B_{2}, E_{1})$ \\ \hline
\end{tabular}%
\end{equation*}

\section{Heun equation and gauged solitons}

Here it will be discussed how the Heun and Whittaker-Hill equations appear
in the construction of both gauged Skyrmions and gauged time-crystals
described above. In the first subsection we will analyze the gauged
Skyrmions and in the second subsection the gauged time-crystals will be
considered. In both subsections, we will use following simple generalization
of the metric (\ref{Minkowski}) 
\begin{equation}
ds^{2}=-dt^{2}+l_{1}^{2}dr^{2}+l_{2}^{2}d\theta ^{2}+l_{3}^{2}d\phi ^{2}\ ,
\label{linearsolution}
\end{equation}%
corresponding to a box with sides with different sizes while the periods of
the coordinates will be the same as in Eq. (\ref{period0}). The reduction of
the coupled gauged Skyrme and Maxwell field equations obtained in \cite%
{gaugsk}\ also holds with the above metric.

It is a quite remarkable feature of the present gauged solitons that, in
both families, one can give a complete analytical description of these
(3+1)-dimensional topological objects in a theory (the gauged Skyrme model),
which is far from being integrable, in terms of the solutions of the Heun
and Whittaker-Hill equations (which are well-known example of
quasi-integrable equations \cite{WH1}, \cite{WH2}). Besides the intrinsic
interest of this result, the present framework clearly shows that the
resurgence paradigm is also very effective in the low energy limit of QCD
coupled to electrodynamics. In particular, electromagnetic perturbations of
the gauged Skyrmions satisfy the Mathieu equation which is very well suited
for the resurgence approach in \cite{Demulder:2016mja}, \cite{dunne1}, \cite%
{Dunne:2016qix}.

In view of the results of \cite{Demulder:2016mja}, the present results
disclose an unexpected relation between the (3+1)-dimensional gauged Skyrme
model and the so-called $\eta $-deformed Principal Chiral Models.

\subsection{Heun equation and gauged Skyrmions}

A direct computation shows that, using the line element in Eq. (\ref%
{linearsolution}), the three coupled gauged Skyrme equations (namely, $%
\mathit{E}^{j}=0$, $j=1$, $2$, $3$) in Eq. (\ref{nonlinearsigma1}) 
\begin{equation*}
D^{\mu }\left( R_{\mu }+\frac{\lambda }{4}\left[ R^{\nu },G_{\mu \nu }\right]
\right) =\mathit{E}^{j}t_{j}=0
\end{equation*}%
and the corresponding four Maxwell equations in Eq. (\ref{maxwellskyrme1})
are greatly simplified by the Skyrme ansatz in Eqs. (\ref{euan}) and (\ref%
{eultc}) and the gauge potential in Eq. (\ref{EMpotans1}).

Indeed, Eq. (\ref{nonlinearsigma1}) reduce to only one Skyrme field equation
(since the third Skyrme equation is identically satisfied while the first
and the second are proportional):%
\begin{eqnarray*}
\mathit{E}^{3} &=&0\ , \\
\mathit{E}^{1} &=&I_{1}P\left[ H\right] \ ,\ \mathit{E}^{2}=I_{2}P\left[ H%
\right] \ ,\ \ I_{1}\neq 0\ ,\ I_{2}\neq 0\ ,
\end{eqnarray*}%
where $I_{j}$ are real and non-vanishing, while 
\begin{align*}
0& =P\left[ H\right] =4\left( X_{1}\sin ^{2}(H)+\frac{\lambda l_{1}^{2}}{2}%
\left( \frac{p^{2}}{l_{2}^{2}}+\frac{q^{2}}{l_{3}^{2}}\right)
+2l_{1}^{2}\right) H^{\prime \prime }+2X_{1}\sin (2H)H^{\prime 2}+4\sin
^{2}(H)X_{1}^{\prime }H^{\prime } \\
& +\left( 2\lambda l_{1}^{4}\left( \frac{pb_{2}}{l_{2}^{2}}+\frac{qb_{3}}{%
l_{3}^{2}}\right) \left( \frac{pb_{2}}{l_{2}^{2}}+\frac{qb_{3}}{l_{3}^{2}}+%
\frac{1}{2}\left( \frac{p^{2}}{l_{2}^{2}}-\frac{q^{2}}{l_{3}^{2}}\right)
\right) -\frac{l_{1}^{4}}{l_{2}^{2}l_{3}^{2}}\frac{\lambda p^{2}q^{2}}{2}-%
\frac{l_{1}^{2}}{4}\left( \frac{p^{2}}{l_{2}^{2}}+\frac{q^{2}}{l_{3}^{2}}%
\right) X_{1}\right) \sin (4H) \\
& -\frac{2l_{1}^{2}}{\lambda }X_{1}\sin (2H)\ ,
\end{align*}%
where 
\begin{equation*}
X_{1}=4\lambda \left( -2l_{1}^{2}b_{1}^{2}+\frac{l_{1}^{2}}{l_{2}^{2}}%
b_{2}(2b_{2}+p)+\frac{l_{1}^{2}}{l_{3}^{2}}b_{3}(2b_{3}-q)\right) \ .
\end{equation*}%
On the other hand, the Maxwell equations reduce to 
\begin{equation*}
b_{I}^{\prime \prime }=-\frac{K}{2}\left( M_{IJ}b_{J}+N_{I}\right) \ ,
\end{equation*}%
with 
\begin{align*}
M_{11}& =4\sin ^{2}(H)\left( 2\lambda H^{\prime 2}+\frac{\lambda l_{1}^{2}}{2%
}\left( \frac{p^{2}}{l_{2}^{2}}+\frac{q^{2}}{l_{3}^{2}}\right) \cos
^{2}(H)+2l_{1}^{2}\right) \ ,\qquad M_{23}=-\frac{l_{1}^{2}}{2l_{3}^{2}}%
\lambda pq\sin ^{2}(2H)\ , \\
M_{22}& =M_{11}+\frac{p}{q}M_{32}\ ,\qquad M_{32}=\frac{l_{3}^{2}}{l_{2}^{2}}%
M_{23}\ ,\qquad M_{33}=M_{11}+\frac{q}{p}M_{23}\ ,\qquad N_{1}=0\ , \\
N_{2}& =\frac{p}{4}M_{11}+\frac{1}{4q}\left( \frac{l_{3}^{2}p^{2}}{l_{2}^{2}}%
-q^{2}\right) M_{23}\ ,\qquad N_{3}=-\frac{q}{4}M_{11}+\frac{1}{4p}\left( 
\frac{l_{3}^{2}p^{2}}{l_{2}^{2}}-q^{2}\right) M_{23}\ .
\end{align*}

Once again, when algebraic relations below hold 
\begin{equation}
X_{1}=-\frac{\lambda l_{1}^{2}}{2}\Big(\frac{p^{2}}{l_{2}^{2}}+\frac{q^{2}}{%
l_{3}^{2}}\Big)=\text{constant}\ ,\qquad \frac{p}{l_{2}^{2}}b_{2}+\frac{q}{%
l_{3}^{2}}b_{3}=-\frac{1}{4}\Big(\frac{p^{2}}{l_{2}^{2}}-\frac{q^{2}}{%
l_{3}^{2}}\Big)\ ,  \label{simpcond}
\end{equation}%
the system of seven coupled non-linear field equations of the gauged Skyrme
model reduce to 
\begin{eqnarray}
&&\Big(8\big(\frac{p^{2}}{l_{2}^{2}}+\frac{q^{2}}{l_{3}^{2}}\big)%
^{-1}+2\lambda \cos ^{2}(H)\Big)H^{\prime \prime }+\sin
(2H)(l_{1}^{2}-\lambda H^{\prime 2})=0\ ,  \label{skyrmesimple} \\
&&b_{3}^{\prime \prime }-\frac{K}{4}(q-4b_{3})\sin ^{2}(H)\Big\{%
4l_{1}^{2}+4\lambda H^{\prime 2}+\lambda l_{1}^{2}\big(\frac{p^{2}}{l_{2}^{2}%
}+\frac{q^{2}}{l_{3}^{2}}\big)\cos ^{2}(H)\Big\}=0\ .  \label{maxwell}
\end{eqnarray}

The above system is a slight generalization of the one obtained in \cite%
{gaugsk}.

The simplest topologically non-trivial solution for the profile $H(r)$ is
given by 
\begin{equation}
H(r)=\frac{l_{1}}{\sqrt{\lambda }}r+h_{0}\ ,  \label{linear}
\end{equation}%
where $h_{0}$ is a constant. It is important to satisfy the necessary
conditions needed to have a non-vanishing topological charge:$\ $%
\begin{equation*}
H(2\pi )=\pi /2,\ H(0)=0 \ , \qquad \text{or} \qquad H(2\pi )=0,\ H(0)=\pi
/2\ .
\end{equation*}

The above conditions fix $l_{1}$ and $h_{0}$ as follows: 
\begin{equation}
l_{1}=\frac{\sqrt{\lambda }}{4}\ ,\ h_{0}=0\ ,\qquad \text{or}\qquad l_{1}=-%
\frac{\sqrt{\lambda }}{4}\ ,\ h_{0}=\frac{\pi }{2}\ .  \label{bctc1}
\end{equation}

Moreover, the energy density of the system is found to be

\begin{align}
\varepsilon & =\frac{K }{2}\left[ \frac{1}{\lambda }+\frac{1}{2}\left( \frac{%
p^{2}}{l_{2}^{2}}+\frac{q^{2}}{l_{3}^{2}}\right) +\left( \frac{1}{\lambda
l_{1}^{2}}X_{1}+16b_{1}^{2}\right) \sin {H}^{2}+\frac{\lambda }{4}\left(
b_{1}^{2}\left( \frac{p^{2}}{l_{2}^{2}}+\frac{q^{2}}{l_{3}^{2}}\right) +%
\frac{1}{4l_{2}^{2}l_{3}^{2}}\left( 2qb_{2}-2pb_{3}+pq\right) ^{2}\right)
\sin {(2H)}^{2}\right]  \notag \\
& \qquad+\frac{1}{2l_{1}^{2}}\left( b_{1}^{\prime 2}+\frac{b_{2}^{\prime 2}}{%
l_{2}^{2}}+\frac{b_{3}^{\prime 2}}{l_{3}^{2}}\right) \ .
\end{align}%
The above expression will be discussed in more details after constructing
the complete solution of the problem in terms of the Heun functions.

In what follows, we will study the equation (\ref{maxwell}). By introducing
the new variables $x$ and $y$ given by 
\begin{equation}
x=\frac{l_{1}}{\sqrt{\lambda }}r+h_{0}\ ,\qquad y=q-4b_{3}\ ,
\label{newcoord}
\end{equation}%
we can rewrite the equation (\ref{maxwell}) as 
\begin{equation}
\frac{d^{2}y}{dx^{2}}+\Big(8K\lambda \ \sin ^{2}x+\Gamma ^{2}\sin ^{2}2x\Big)%
y=0\ ,  \label{eqy}
\end{equation}%
with a non-negative constant $\Gamma \geq 0$, 
\begin{equation*}
\Gamma ^{2}:=\frac{K\lambda ^{2}}{4}\Big(\frac{p^{2}}{l_{2}^{2}}+\frac{q^{2}%
}{l_{3}^{2}}\Big)\ .
\end{equation*}%
The equation (\ref{eqy}) can be cast into the famous confluent Heun's
equation, 
\begin{equation}
\frac{d^{2}}{dz^{2}}Y(z)+\Big(\frac{\gamma }{z}+\frac{\delta }{z-1}+\epsilon %
\Big)\frac{d}{dz}Y(z)+\frac{\alpha z-q}{z(z-1)}Y(z)=0\ 
\end{equation}%
where 
\begin{eqnarray}
&&z=\cos ^{2}x\ ,\qquad Y(z)=e^{-i\Gamma z}\ y(\arccos \sqrt{z}) \ ,  \notag
\\
&&\gamma =\delta =1/2\ ,\qquad \epsilon =2i\Gamma \ ,\qquad \alpha =i\Gamma
+2K\lambda \ ,\qquad q=i\Gamma /2+2K\lambda \ .
\end{eqnarray}%
A general solution to this equation is found to be 
\begin{eqnarray}
&&Y(z)=C_{1}\ \text{HeunC}(i\Gamma+2K\lambda, 1/2, 1/2, 2i\Gamma , i\Gamma
/2+2K\lambda ;z)  \notag \\
&&\hspace{0.5in}+C_{2}\sqrt{z}\ \text{HeunC}(2i\Gamma+2K\lambda, 3/2, 1/2,
2i\Gamma , 3i\Gamma /2+2K\lambda - 1/4;z)\ ,
\end{eqnarray}%
where $C_{1}$ and $C_{2}$ are integration constants, and $\text{HeunC}%
(\alpha, \gamma, \delta, \epsilon, q ;z)$ is the confluent Heun's function.

\subsubsection{Electric field, magnetic field and boundary conditions}

The non-zero components of the electromagnetic tensor in our configuration
are given by 
\begin{equation*}
E_{r}=\frac{\sqrt{l_{3}^{2}p^{2}+l_{2}^{2}q^{2}}}{l_{3}^{2}p}b_{3}^{\prime
}\ ,\quad B_{\theta }=\frac{1}{l_{1}^{2}l_{3}^{2}}b_{3}^{\prime }\ ,\quad
B_{\phi }=\frac{q}{l_{1}^{2}l_{3}^{2}p}b_{3}^{\prime }\ .
\end{equation*}%
The requeriment that the electric field on the surface of the box ($r=0$ and 
$r=2\pi $) vanishes leads to $b_{3}^{\prime }(0)=b_{3}^{\prime }(2\pi )=0$.
Since in Eq. (\ref{maxwell}) the unknown is $b_{3}(r)$, imposing a condition
on the electric and magnetic field induces Neumann boundary conditions for
this potential (note also that since $b_{3}$ is part of a gauge conection,
it is defined up to an additive constant). Notwithstanding Eq. (\ref{maxwell}%
) can be solved analytically in terms of Confluent Heun functions, it is
simpler to impose the boundary condition in a numerical integration. Then
Eq. (\ref{eqy}) defines the following Sturm-Liouville problem%
\begin{equation}
\frac{d}{dx}\left( \hat{p}\left( x\right) \frac{d}{dx}y\left( x\right)
\right) +\hat{Q}\left( x\right) y\left( x\right) =-\Gamma ^{2}\hat{w}\left(
x\right) y\left( x\right) \ ,  \label{sturmliouville}
\end{equation}%
where 
\begin{eqnarray*}
\hat{w}\left( x\right) &=&\sin ^{2}2x\ ,\ \ \hat{p}\left( x\right) =1 \\
\hat{Q}\left( x\right) &=&8K\lambda \sin ^{2}x\ .
\end{eqnarray*}%
Since $\Gamma ^{2}$ plays the role of the eigenvalue of the Sturm-Liouville
problem defined above, then there is a countable infinity of values for $%
\Gamma $ which are consistent with the boundary conditions imposed. Since $p$
and $q$ are integers, the quantization of $\Gamma $ induces a quantization
on the possible values of the volume within which the gauged baryons are
confined!

\begin{figure}[h]
\centering
\includegraphics[scale=0.58]{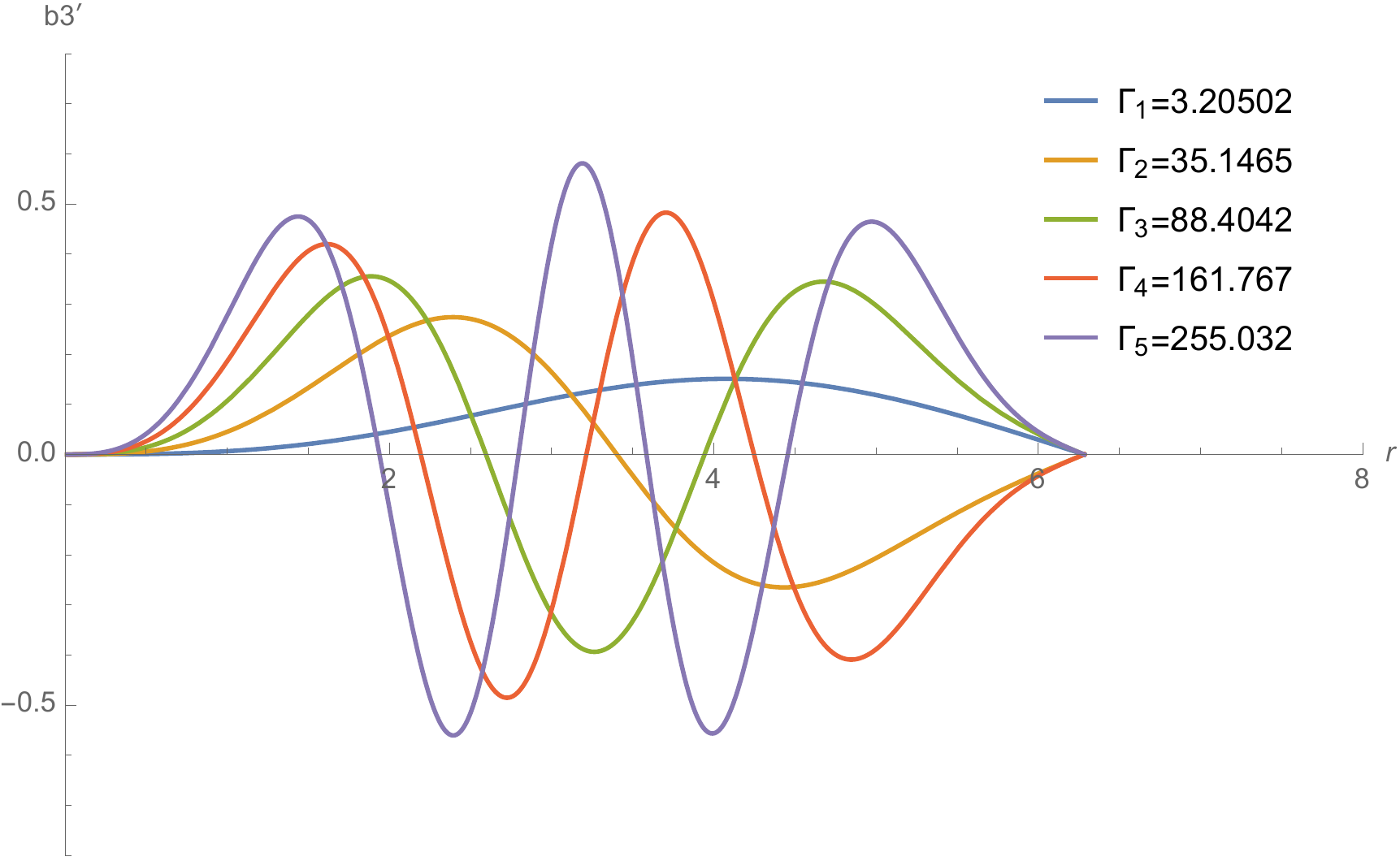}
\caption{Behaviour of $b_3^{\prime }(r)$ for the first values of $\Gamma$.}
\label{db3}
\end{figure}

Figure \ref{db3} shows the profiles for $b_3^{\prime }$ for the first five
allowed values of $\Gamma$. The non-triviality of the profile inside the box
is due to the presence of the current.

\subsubsection{Gauged Skyrmion energy}

\begin{figure}[h]
\centering
\includegraphics[scale=0.7]{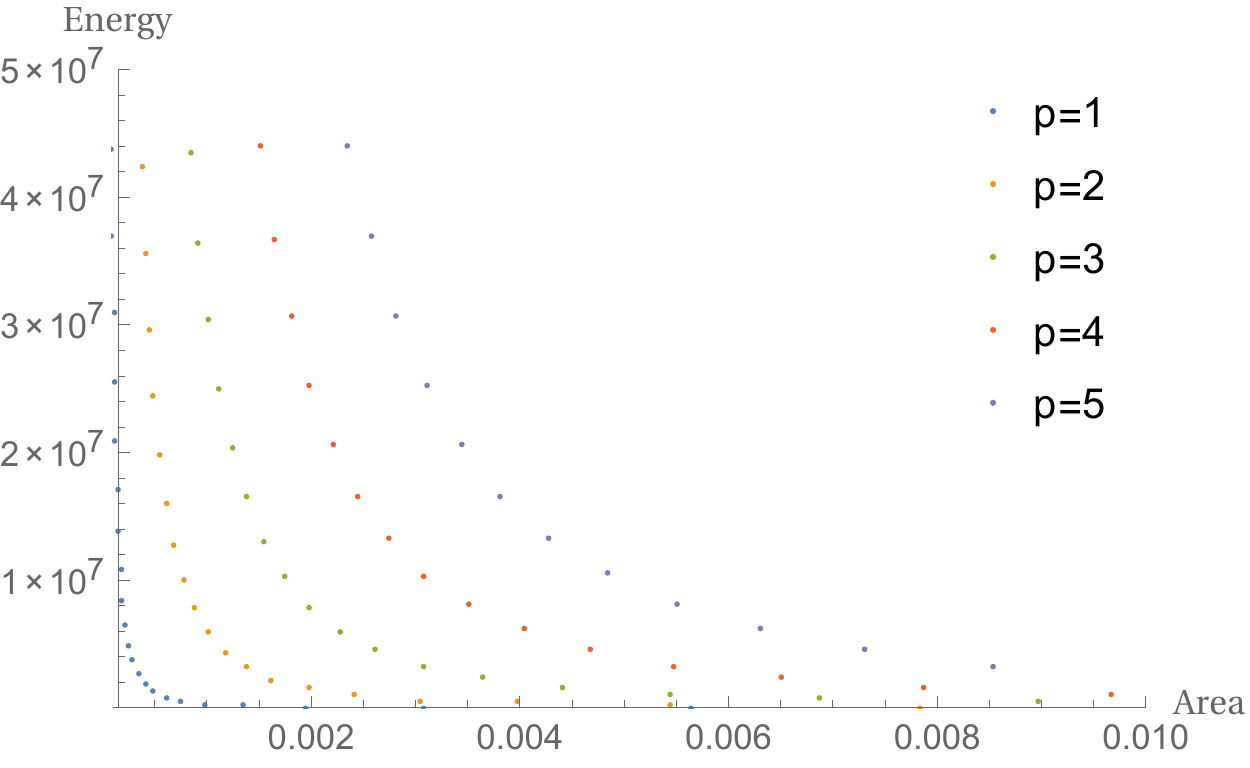}
\caption{Energy of the system as a function of the area for $p=1,2,3,4,5$.}
\label{energyplot}
\end{figure}

In order to do the energy plots it is enough to consider the case in which

\begin{equation*}
p=q\ ,\ \quad l_{2}=l_{3}\ ,
\end{equation*}%
so that the Baryon charge is%
\begin{equation*}
B=p^{2}=q^{2}\ ,
\end{equation*}%
while the area $\widetilde{A}$ of the box orthogonal to the $r$-axis is%
\begin{equation*}
A=\frac{\widetilde{A}}{8\pi ^{2}}=l_{2}^{2}=l_{3}^{2}\ .
\end{equation*}%
On the other hand, the total volume is%
\begin{equation*}
V=2\pi l_{1}\widetilde{A}=32\pi ^{3}\frac{\sqrt{\lambda }}{4}l_{2}^{2}\ .
\end{equation*}
Thus, the energy density of the system reads 
\begin{align}
\varepsilon = & \frac{ K }{2}\left[ \frac{1}{\lambda }+\frac{B}{A}+\left( 
\frac{1}{\lambda l_{1}^{2}}X_{1}+16b_{1}^{2}\right) \sin {H}^{2}+\frac{%
\lambda }{4}\left( 2\frac{B}{A}b_{1}^{2}+\frac{1}{4A^{2}}\left(
2B^{1/2}b_{2}-2B^{1/2}b_{3}+B\right) ^{2}\right) \sin {(2H)}^{2}\right] 
\notag \\
& +\frac{1}{2l_{1}^{2}}\left( b_{1}^{\prime 2}+\frac{b_{2}^{\prime
2}+b_{3}^{\prime 2}}{A}\right)  \notag \\
= & \frac{K}{32 A^2 \lambda} ( 16A^2+24 AB\lambda +B^2\lambda^2-B\lambda
(8A\cos(2H)+B\lambda \cos(4H))  \notag \\
& -32 K\lambda (B^{1/2}-2b_3)b_3( 8A+B\lambda\cos(2H))\sin^2(H)+256 A
b_3^{\prime 2 }) \ .
\end{align}%
Figure \ref{energyplot} show the energy as function of the area. The
divergence for low values of the area is expected on general ground since,
at very small distances, the Skyrme model should be replaced by QCD.

\subsubsection{Relation with the Whittaker-Hill equation}

If we take a coordinate transform 
\begin{equation}
y=q-4b_{3}\ ,\qquad x=\frac{l_{1}}{\sqrt{\lambda }}r+h_{0}\ ,
\end{equation}%
and use the equation (\ref{linear}) in (\ref{maxwell}), then we have a
Whittaker-Hill equation 
\begin{equation}
\frac{d^{2}y}{dx^{2}}+\left( 4\alpha s\cos (2x)+2\alpha ^{2}\cos
(4x)+\lambda _{0}\right) y=0\ ,  \label{WHeq}
\end{equation}%
where 
\begin{eqnarray}
\alpha ^{2} &=&-\frac{K\lambda ^{2}}{16}\Big(\frac{p^{2}}{l_{2}^{2}}+\frac{%
q^{2}}{l_{3}^{2}}\Big)\ ,\qquad s=4i\Big(\frac{p^{2}}{Kl_{2}^{2}}+\frac{q^{2}%
}{Kl_{3}^{2}}\Big)^{-1/2}\ , \\
\qquad \lambda _{0} &=&K\lambda \Big\{4+\frac{\lambda }{8}\Big(\frac{p^{2}}{%
l_{2}^{2}}+\frac{q^{2}}{l_{3}^{2}}\Big)\Big\}\ ,
\end{eqnarray}%
and $\lambda _{0}$ is an eigenvalue of the differential operator. It is not
true, however, that the Heun equation discussed in the previous subsection
is equivalent to the Whittaker-Hill equation (indeed, the relation of the
present gauged solitons with the Heun equation is more natural), because the
parameters $\alpha $ and $s$ are not independent (unlike what would happen
in a ``proper" Whittaker-Hill equation). Thus, $s$ need not be an integer.
Indeed, 
\begin{equation}
\alpha =-K\lambda /s\ ,\qquad \lambda _{0}=4K\lambda -2K^{2}\lambda
^{2}/s^{2}\ .
\end{equation}

Now, we can compare Eq. (\ref{WHeq}) with the Whittaker-Hill's equation in 
\cite{Demulder:2016mja}, 
\begin{equation}
\psi ^{\prime \prime }+(a-2b\cos 2x-2c\cos 4x)\psi =0\ ,  \label{wheq}
\end{equation}%
where 
\begin{equation}
a=\lambda _{0}\ ,\qquad b=-2\alpha s\ ,\qquad c=-\alpha ^{2}\ .
\end{equation}%
From their analysis we can obtained the resurgent parameter of our model,
that is 
\begin{equation}
g^{2}=\frac{1}{2\sqrt{4\alpha ^{2}+2\alpha s}}=\frac{il_{2}l_{3}}{\sqrt{%
K\lambda \left( \lambda l_{2}^{2}q^{2}+\lambda
l_{3}^{2}p^{2}+8l_{2}^{2}l_{3}^{2}\right) }}\ .
\end{equation}

\subsubsection{Perturbations, Mathieu equation and resurgence}

In this subsection we will discuss how typical electromagnetic perturbations
of the gauged Skyrmions constructed above disclose the resurgence structure
of these (3+1)-dimensional gauged solitons. Before entering into the
technical details, it is worth to remind how standard large \textbf{N}
arguments can simplify the analysis of the present subsection (see for a
detailed review chapter 4-in particular, section 4.2-of the classic
reference \cite{skyrev0}). As it is well known, in the leading 't Hooft
approximation, in meson-Baryon scattering, the very heavy Baryon (the
Skyrmion in our case) is essentially unaffected and, basically, only the
meson can react. This is even more so in the photon-Baryon semiclassical
interactions (due to the masslessness of the photon). Thus, in this
approximation, electromagnetic perturbations perceive the Skyrmions as an
effective medium. From the practical point of view, this simplifies the
analysis since one can neglect the perturbations of the Skyrmions
(suppressed by powers of 1/\textbf{N}) and one is allowed to only consider
the reaction of the Maxwell equations to perturbations around the gauged
Skyrmion background. In other words, one can consider electromagnetic
perturbations of Eqs. (\ref{maxwellskyrme1}) and (\ref{current}) in which
the background solution is the gauged Skyrmion defined in Eqs. (\ref{ans1}),
(\ref{EMpotans1}), (\ref{linearsolution}), (\ref{simpcond}), (\ref{linear})
and (\ref{bctc1}).

As it is well known, the full power of resurgence manifests itself
especially in relating the perturbative expansion around the trivial vacuum
with the perturbative expansions around non-trivial saddles. In the present
case, the analysis of the full perturbative expansion around the gauged
solitons constructed in the previous sections would correspond to the
analysis of seven coupled linear PDEs in the background of gauged solitons
discussed above. This analysis is extremely difficult even numerically.
Consequently, we considered a simpler (yet interesting) situation in which
the Skyrme background is considered to be fixed and one analyzes magnetic
perturbations of the Maxwell equations in the background of the gauged
Skyrmion itself (this situation is enough to show that resurgence appears
also in the gauged Skyrme model).

\begin{figure}[h]
\centering
\includegraphics[scale=0.6]{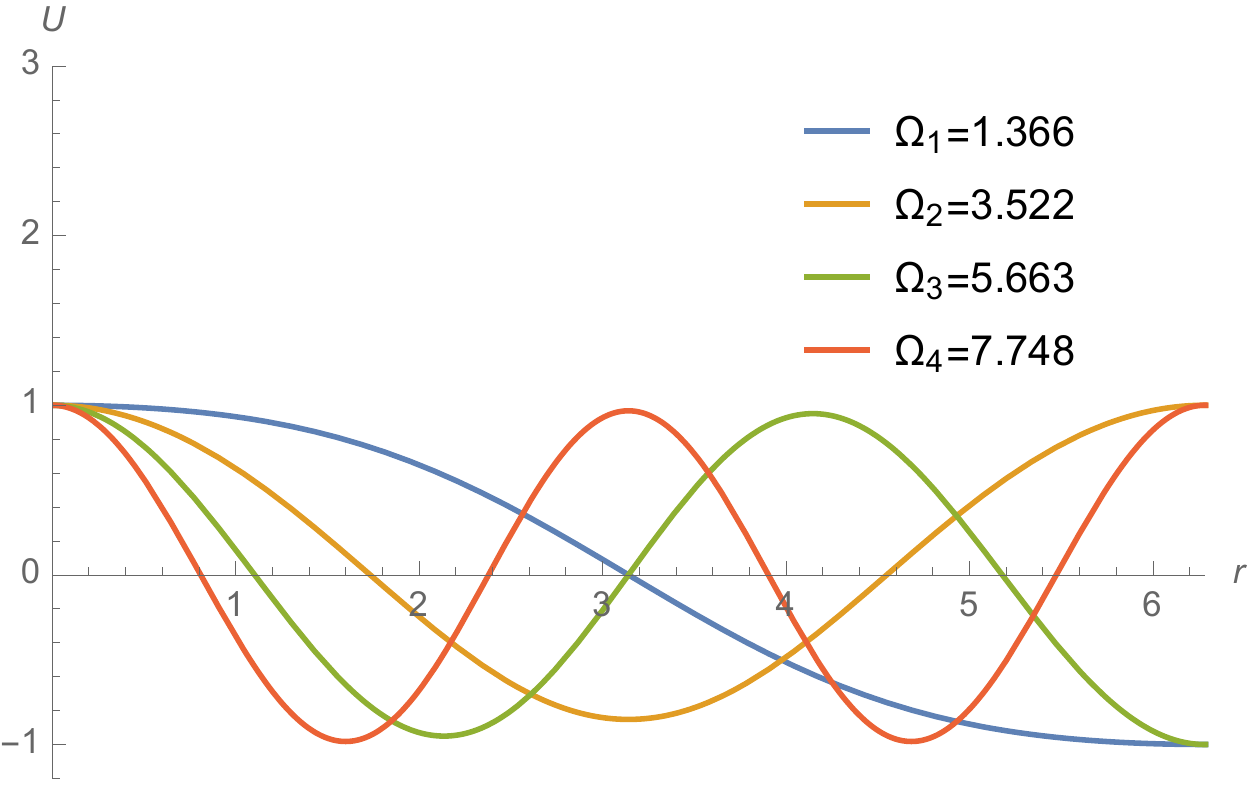}\qquad \includegraphics[scale=0.6]{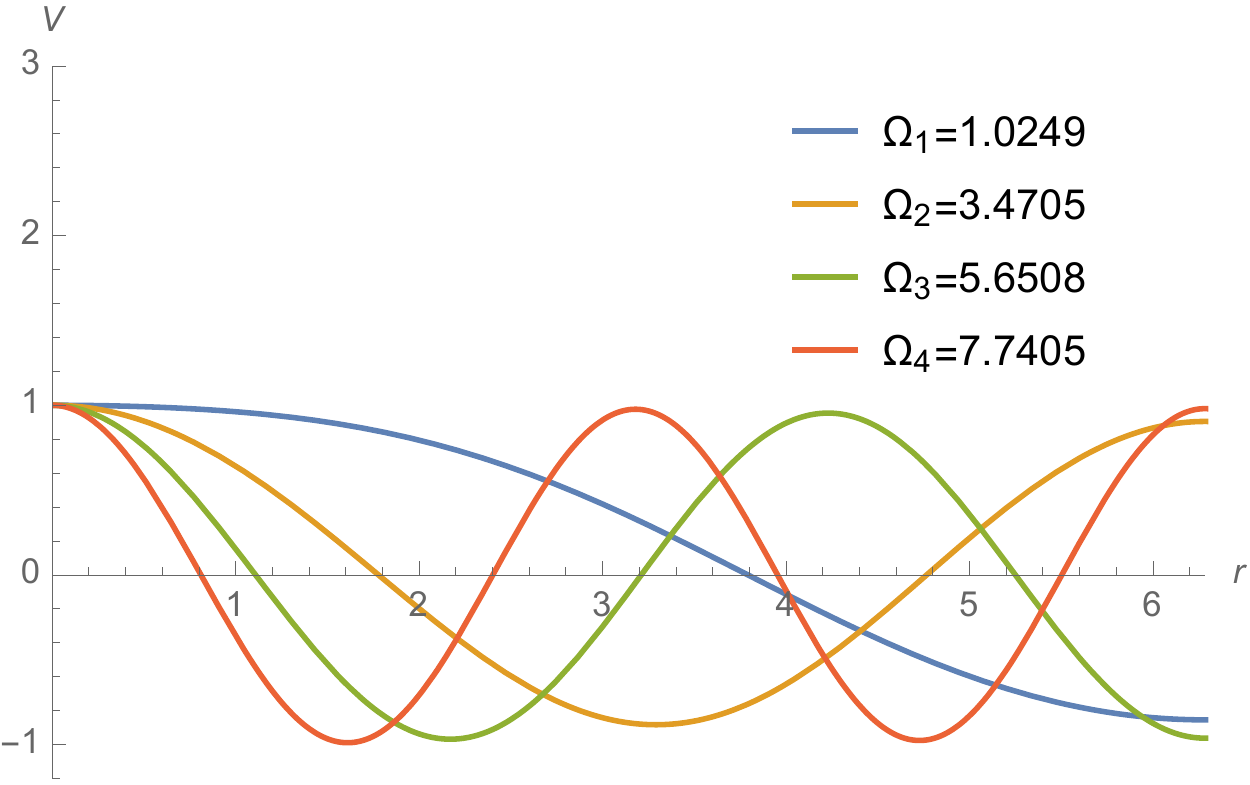}
\caption{First four normal modes for the normal coordinates $U$ and $V$,
together with the corresponding normal frequencies. Neumann boundary
conditions have been imposed on the electromagnetic perturbation.}
\label{Pert1}
\end{figure}

Thus, let's consider the following perturbations around the solutions
defined in Eqs. (\ref{ans1}), (\ref{EMpotans1}), (\ref{simpcond}), (\ref%
{linear}) and (\ref{bctc1}): 
\begin{equation}
b_{2}(r)\rightarrow b_{2}(r)+\epsilon c_{2}(r)\sin (\Omega t)\ ,\quad
b_{3}(r)\rightarrow b_{3}(r)+\epsilon c_{3}(r)\sin (\Omega t)\ ,
\label{defpert}
\end{equation}%
$\Omega $ being the frequency of the perturbation: the mathematical problem
is to find how $\Omega $\ depends on the parameters of the problem. To first
order in $\epsilon $, Eqs. (\ref{maxwellskyrme1}) and (\ref{current}) reduce
to 
\begin{align}
0 \quad = \quad & 4l_{3}^{2}c_{2}^{\prime \prime }(r)-Kl_{1}^{2}pq\lambda
c_{3}(r)\sin ^{2}(2H)  \notag \\
& +\frac{1}{2}l_{1}^{2}c_{2}(r)\left( 32Kl_{3}^{2}+Kq^{2}\lambda
+8l_{3}^{2}\Omega ^{2}-32Kl_{3}^{2}\cos (2H)-Kq^{2}\lambda \cos (4H)\right)
\ ,  \label{persystem1} \\
0 \quad = \quad & 4l_{2}^{2}c_{3}^{\prime \prime }(r)-Kl_{1}^{2}pq\lambda
c_{2}(r)\sin ^{2}(2H)  \notag \\
& +\frac{1}{2}l_{1}^{2}c_{3}(r)\left( 32Kl_{2}^{2}+Kp^{2}\lambda
+8l_{2}^{2}\Omega ^{2}-32Kl_{2}^{2}\cos (2H)-Kp^{2}\lambda \cos (4H)\right)
\ .  \label{persystem2}
\end{align}%
In the present subsection, we will consider 
\begin{equation*}
l_{3}=l_{2}\ ,\quad p=q\ ,
\end{equation*}%
and introduce the normal variables 
\begin{equation*}
U(r)=c_{3}(r)+c_{2}(r),\quad V(r)=c_{3}(r)-c_{2}(r) \ ,
\end{equation*}%
such that the system given by Eq. (\ref{persystem1}) and Eq. (\ref%
{persystem2}) decouples and leads to 
\begin{eqnarray}
\frac{d^2U}{dr^2}+l_{1}^{2}\left( 8K\sin ^{2}(H)+\Omega_U ^{2}\right) U=0\ ,
\label{dunnecorr1} \\
\frac{d^2V}{dr^2}+ l_1^2\left( \frac{Kq^2\lambda}{2l_2^2} \sin^2(2 H) +
8K\sin^2(H) +\Omega_V^2 \right) V =0 \ .  \label{dunnecorr2F}
\end{eqnarray}%
These equations correspond to a \textit{Mathieu equation} and a
Whittaker-Hill equation, respectively (see \cite{Dunne:2016qix} for a
resurgence analysis of the Mathieu equation). It is interesting to note that
the equation for the normal coordinate $U$ does not depend on the details of
the electromagnetic background defined by $b_3(r)$, while the equation for $V
$ depends explicitly on the quotient $q^2/l_2^2$ which is different for each
of the possible background configurations and depends on the number of nodes
of the function $b_3(r)$ within the cavity. We will focus on perturbations
of the groundstate, i.e. the node-less $b^{\prime }_3$. We have introduced
an index for the normal frequencies $\Omega_U,\ \Omega_V$ associated with
the normal coordinates $U(r),\ V(r)$, respectively. It is natural to
restrict the perturbations $c_i$ to fulfil the same boundary conditions than
the unperturbed solution, therefore $c^{\prime }_i(r=0)=c^{\prime
}_i(r=2\pi)=0$. This induces a Neumann boundary condition for $U$ and $V$,
such that one has to solve Eqs. (\ref{dunnecorr1}) and (\ref{dunnecorr2F})
restricted to 
\begin{align}
U^{\prime }(r=0)=U^{\prime }(r=2\pi)=0 \ .  \label{bcpert} \\
V^{\prime }(r=0)=V^{\prime }(r=2\pi)=0 \ .  \label{bcpert2}
\end{align}
As expected, this quantizes the normal frequencies of the perturbations $%
\Omega_U$ and $\Omega_V$ which leads to the normal modes of the system
inside the box. Thus, the interesting problem is to determine $%
\Omega_{(U,V)} ^{2}=\Omega_{(U,V)}^{2}\left( K,\lambda ;n\right)$, with $n$
an integer labelling the mode. Namely, we would like to know how the
frequency of the electromagnetic perturbation depends on the label $n$ and
on the coupling constants $K$ and $\lambda $ of the theory. Obviously, since
the problem is linear, the general solutions will be given by an arbitrary
linear superposition of the normal modes multiplied by harmonic time factors
with the corresponding normal frequencies. Figure \ref{Pert1} shows the
first four normal modes for the normal coordinates $U$ and $V$ 

\begin{figure}[h]
\centering
\includegraphics[scale=0.27]{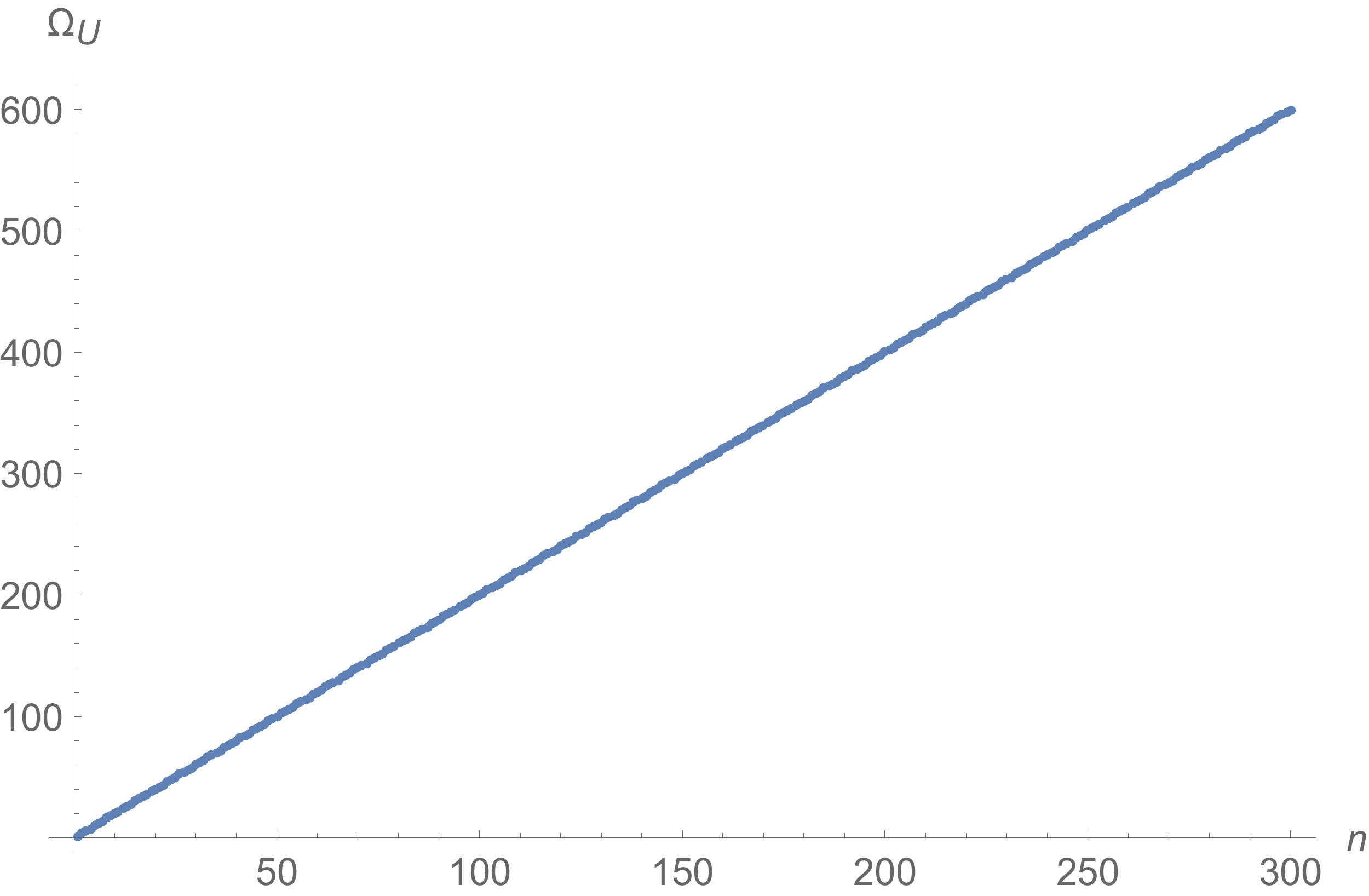}\quad %
\includegraphics[scale=0.18]{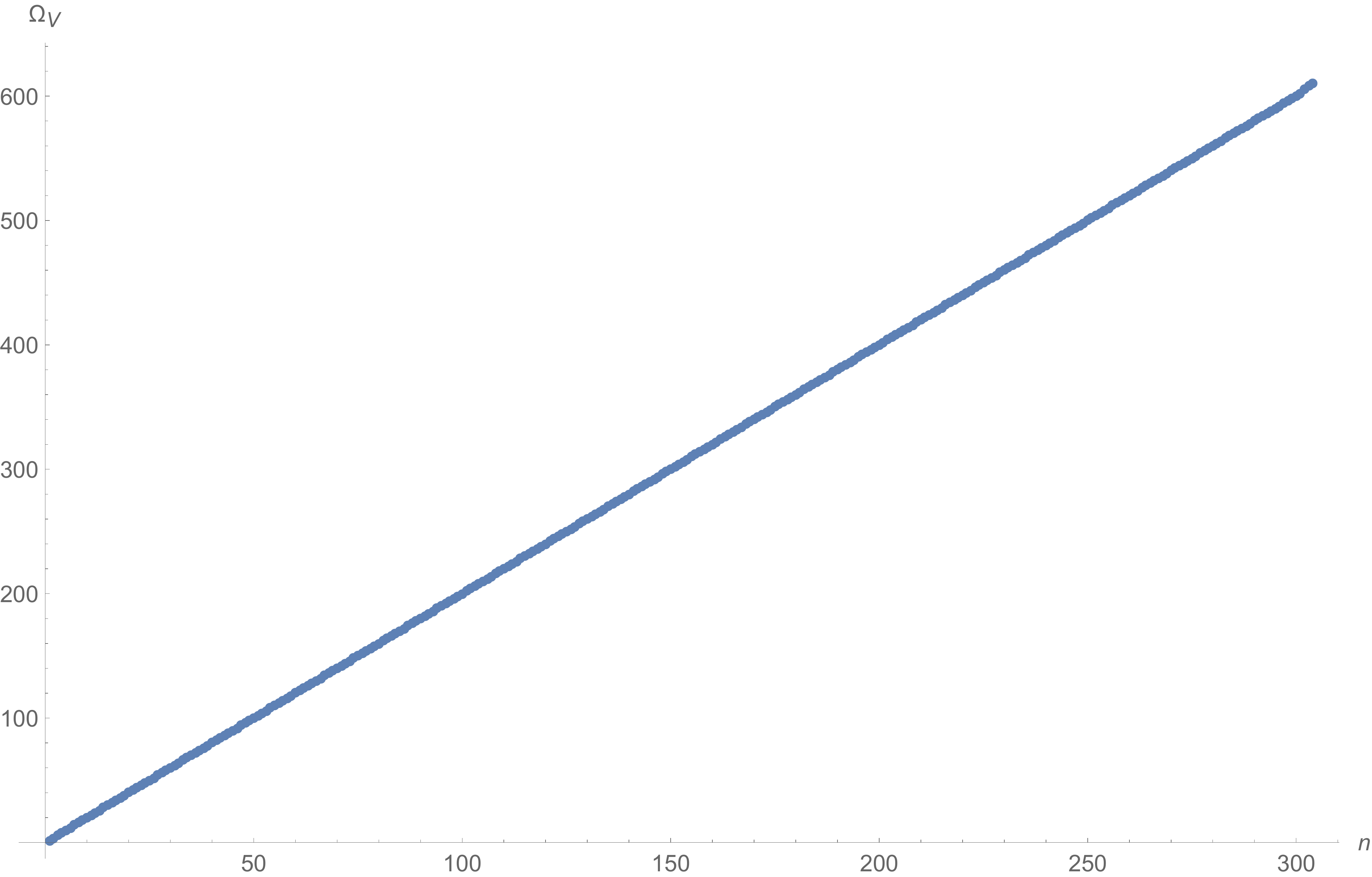}
\caption{The spectrum of the electromagnetic perturbation. One sees that
rapidly, as a function of the mode number, the frequencies tend to an
equispaced, i.e. linear spectrum. The equation for the normal variable $U$
depends on the details of the background configuration. We have selected the
nodeless configuration for $b_3^{\prime }(r)$ as a background.}
\label{Un}
\end{figure}

The Eq. (\ref{dunnecorr1}) can be brought into the standard Mathieu form 
\begin{equation}
U^{\prime \prime }+(A-2Q\cos (2x))U=0\ ,  \label{dunnecorr1x}
\end{equation}%
when the parameters are related as 
\begin{equation}
A=\left( 4K+\Omega_U ^{2}\right) l_{1}^{2} \ , \quad 2Q=4Kl_{1}^{2}\ .
\label{dunnecorr2}
\end{equation}%
The comparison with Eqs. (7) and (8) of \cite{Dunne:2016qix} shows the
correspondence between the Skyrme and Mathieu parameters: 
\begin{eqnarray}
\frac{2}{\hbar ^{2}} &=&Kl_{1}^{2}\ , \quad \left( 4K+\Omega_U ^{2}\right)
l_{1}^{2}=\frac{8u}{\hbar ^{2}}\ \Rightarrow  \label{dunnecorr3} \\
\Omega_U ^{2} &=&\frac{8u}{l_{1}^{2}\hbar ^{2}}-4K\ , \quad \hbar _{eff}^{2}=%
\frac{2}{Kl_{1}^{2}}=\frac{32}{K\lambda }\ ,  \label{dunnecorr4}
\end{eqnarray}%
where the combination $32/K\lambda $ plays the role of the ``effective
Planck constant" $\hbar _{eff}^{2}$ of the problem so that the parameter $u$
does not depend separately on $K$ and $\lambda $ but only on their product
(as well as on the label $n$ of the discrete energy level). Well known
results in the theory of the Mathieu equation can be used to determine the
spectrum (in particular, the parameter $\Omega_U $) of the above
perturbations in Eqs. (\ref{defpert}). Let us focus on Eqs. (\ref{dunnecorr1}%
), (\ref{dunnecorr2}), (\ref{dunnecorr3}) and (\ref{dunnecorr4}). The
results of \cite{Dunne:2016qix} which can be applied directly to our case
are (with the obvious replacement $\hbar \rightarrow \hbar _{eff}$):

1) To expand $u\left( K,\lambda ;n\right) $ (or equivalently $\Omega \left(
K,\lambda ;n\right) ^{2}$ through Eq. (\ref{dunnecorr4})) in power series of
the effective Planck constant $\hbar _{eff}^{2}$\ is not enough to get a
mathematically well-defined answer. The perturbative series is not even
Borel summable. However, the inclusion of non-perturbative contributions
discloses the resurgent phenomenon.

2) One needs to express $u$\ as a trans-series:%
\begin{equation}
\left( u_{trans}\left( \hbar _{eff};n\right) \right) ^{2}=\sum_{k=0}^{\infty
}\sum_{j=0}^{\infty }\sum_{l=1}^{k-1}c_{k,j,l}\left( n\right) \left( \hbar
_{eff}\right) ^{j}\left( \frac{\exp \left[ -\frac{S}{\hbar _{eff}}\right] }{%
\left( \hbar _{eff}\right) ^{n+1/2}}\right) ^{k}\left( \ln \left( -\frac{1}{%
\hbar _{eff}}\right) \right) ^{l} \ ,  \label{trans1}
\end{equation}%
where, with the normalizations in Eqs. (7), (8) and (9) of \cite%
{Dunne:2016qix} the $S$ in the exponential factors (the ``instanton" action)
in the above trans-series is 
\begin{equation*}
S=8\ .
\end{equation*}

3) The above trans-series in Eq. (\ref{trans1}) allows to define clearly a
strong coupling ($n\hbar _{eff}\gg 1$) regime and weak coupling regime ($%
n\hbar _{eff}\ll 1$). The corresponding expansions (reviewed in \cite%
{Dunne:2016qix}) can be applied directly to the present case. We will not
report these expansions here\footnote{%
The reader can refer to \cite{Dunne:2016qix} and references therein.} since
the main aim of the present subsection is to show the explicit relations of
the present gauged Skyrmion and its electromagnetic perturbations with
(well-known results on) the Mathieu equation.

The analogy with the Mathieu equation analyzed in \cite{Dunne:2016qix}\ is
not complete since, in that reference, the Mathieu equation was interpreted
as a Schrodinger equation so that the unknown function in Eq. (7) of \cite%
{Dunne:2016qix} is a complex wave function satisfying the boundary
conditions in Eq. (36) of the same reference. In the present case, the
unknown function $U$ in Eq. (\ref{dunnecorr1}) is real and satisfies Eq. (%
\ref{bcpert}). On the other hand, some of the results in \cite{Dunne:2016qix}
can be applied directly: in particular, the results which do not depend on
the boundary conditions in Eq. (36) of \cite{Dunne:2016qix} (such as the
ones on the asymptotic expansions of $u\left( K,\lambda ;n\right) $) hold in
the present case as well. It is a quite remarkable feature of the present
gauged solitons in (3+1) dimensions that the resurgent structures are so
transparent in this setting. The gauged Skyrme model in (3+1) dimensions is
definitely not a toy model and yet the importance of the resurgence
interplay between the perturbative expansion and the non-perturbative
contributions is manifest. We hope to come back on the appearance of
resurgence in the gauged Skyrme model in a future publication. \newline
On the other hand, the normal coordinate $V(r)$ is determined by equation (%
\ref{dunnecorr2}) which being a Whittaker-Hill equation admits a mapping
with the parameters in Eq. (\ref{WHeq}) to Eq. (\ref{wheq}) by setting 
\begin{align*}
\alpha^2= -\frac{Kq^2\lambda l_1^2}{8 l_2^2} \ , \qquad s= \frac{iKl_1l_2}{q}%
\sqrt{\frac{8}{K\lambda}} \ , \\
\lambda_0 = \frac{l_1^2}{4 l_2^2}\left( 16Kl_2^2+Kq^2\lambda +4 l_2^2
\Omega^2 \right) \ ,
\end{align*}
and 
\begin{align*}
a=\frac{l_1^2}{4 l_2^2}\left( 16Kl_2^2+Kq^2\lambda +4 l_2^2 \Omega^2 \right)
\ , \quad b= 2K l_1^2 \ , \quad c=\frac{K q^2\lambda l_1^2}{8 l_2^2} \ .
\end{align*}
Consequently, the resurgence parameter in this case is 
\begin{align*}
g^2 = \frac{i}{2\sqrt{4c+b}}=\frac{i l_2}{\sqrt{2K\lambda(\lambda q^2+4l_2^2)%
}l_1} \ .
\end{align*}

\subsection{Heun equation and gauged time-crystals}

The Skyrme configuration for time crystals reads 
\begin{equation*}
\alpha =\frac{\phi }{2}\ ,\quad \beta =H(r)\ ,\quad \rho =\frac{\omega
\gamma }{2}\ ,
\end{equation*}%
with the coordinates ordering as 
\begin{equation*}
x^{\mu }=\left( \gamma ,r,t,\phi \right) \ .
\end{equation*}%
Also in this case, a direct computation shows that, using the line element
in Eq. (\ref{linearsolution}), the three coupled gauged Skyrme equations
(namely, $\mathit{E}^{j}=0$, $j=1$, $2$, $3$) in Eq. (\ref{nonlinearsigma1}) 
\begin{equation*}
D^{\mu }\left( R_{\mu }+\frac{\lambda }{4}\left[ R^{\nu },G_{\mu \nu }\right]
\right) =\mathit{E}^{j}t_{j}=0
\end{equation*}%
and the corresponding four Maxwell equations in Eq. (\ref{maxwellskyrme1})
are greatly simplified by the Skyrme ansatz in Eqs. (\ref{euan}), (\ref{cotc}%
), (\ref{eultc}) and the gauge potential in Eq. (\ref{EMpotans1}).

Indeed, Eq. (\ref{nonlinearsigma1}) reduce to only one Skyrme field equation
(since the third Skyrme equation is identically satisfied while the first
and the second are proportional):%
\begin{eqnarray*}
\mathit{E}^{3} &=&0\ , \\
\mathit{E}^{1} &=&I_{1}P_{TC}\left[ H\right] \ ,\ \mathit{E}^{2}=I_{2}P_{TC}%
\left[ H\right] \ ,\ \ I_{1}\neq 0\ ,\ I_{2}\neq 0 \ ,
\end{eqnarray*}%
where $I_{j}$ are real and non-vanishing while the only non-trivial Skyrme
field equation $P_{TC}\left[ H\right] =0$ reads%
\begin{equation}
\begin{split}
& 4\left( l_{3}^{2}\left( 4-\lambda \omega ^{2}\right) +\frac{l_{3}^{2}}{%
l_{1}^{2}}X_{2}\sin ^{2}(H)+\lambda \right) H^{\prime \prime }+\frac{%
2l_{3}^{2}}{l_{1}^{2}}X_{2}\sin (2H)H^{\prime 2}+\frac{4l_{3}^{2}}{l_{1}^{2}}%
\sin ^{2}(H)X_{2}^{\prime }H^{\prime } \\
& +\left[ \frac{1}{4}(l_{3}^{2}\omega ^{2}-1)X_{2}+\frac{\lambda l_{1}^{2}}{%
l_{3}^{2}}(2l_{3}^{2}\omega b_{1}-2b_{3}-1)(2l_{3}^{2}\omega
b_{1}-2b_{3}-l_{3}^{2}\omega ^{2})\right] \sin (4H)-\frac{2l_{3}^{2}}{%
\lambda }X_{2}\sin (2H)=0 \ ,
\end{split}
\label{tcprof}
\end{equation}%
where 
\begin{equation}
X_{2}(r)=8\lambda \Big(l_{1}^{2}b_{1}(\omega -2b_{1})+\frac{2l_{1}^{2}}{%
l_{2}^{2}}b_{2}^{2}+\frac{l_{1}^{2}}{l_{3}^{2}}b_{3}\left( 1+2b_{3}\right) %
\Big)\ .
\end{equation}%
The Maxwell equations are written in the same form as the previous section,
where the matrix for this case is given by 
\begin{align*}
& M_{11}=2\sin ^{2}(H(r))\left( 4\lambda H^{\prime 2}+\frac{\lambda l_{1}^{2}%
}{l_{3}^{2}}\cos ^{2}(H)+4l_{1}^{2}\right) \ , \\
& M_{13}=-\frac{\lambda \omega l_{1}^{2}}{2l_{3}^{2}}\sin ^{2}(2H) \ , \\
& M_{22}=M_{11}+l_{3}^{2}\omega M_{13} \ , \\
& M_{33}=M_{11}+\frac{l_{3}^{2}\omega ^{2}+1}{\omega }M_{13} \ , \\
& M_{31}=-l_{3}^{2}M_{13} \ ,
\end{align*}%
while 
\begin{equation*}
N=\left( \frac{1}{4}(M_{13}-\omega M_{11}),0,\frac{1}{4}\left( \frac{\left(
2l_{3}^{2}\omega ^{2}+1\right) }{\omega }M_{13}+M_{11}\right) \right) .
\end{equation*}
\newline

When we impose the relations%
\begin{equation}
X_{2}=\frac{\lambda l_{1}^{2}}{l_{3}^{2}}\left( l_{3}^{2}\omega
^{2}-1\right) =\text{constant}\ ,\qquad b_{3}(r)=l_{3}^{2}\omega b_{1}(r)-%
\frac{l_{3}^{2}\omega ^{2}}{4}-\frac{1}{4}\ ,
\end{equation}%
the field equations are reduced to 
\begin{equation}
2\big(\lambda (l_{3}^{2}\omega ^{2}-1)\cos ^{2}(H)-4l_{3}^{2}\big)H^{\prime
\prime }+\left( l_{3}^{2}\omega ^{2}-1\right) \sin (2H)\left(
l_{1}^{2}-\lambda H^{\prime 2}\right) =0\ ,  \label{skeqcrystal}
\end{equation}%
\begin{equation}
b_{1}^{\prime \prime }+\frac{K}{8}(\omega -4b_{1})\sin ^{2}(H)\Big(%
l_{1}^{2}\left( \lambda \omega ^{2}-8\right) -\frac{l_{1}^{2}}{l_{3}^{2}}%
\lambda +\frac{\lambda l_{1}^{2}}{l_{3}^{2}}(\omega ^{2}l_{3}^{2}-1)\cos
(2H)-8\lambda H^{\prime 2}\Big)=0\ .  \label{mxeqcrystal}
\end{equation}%
Eq. (\ref{skeqcrystal}) has the solution 
\begin{equation*}
H(r)=\frac{l_{1}}{\sqrt{\lambda }}r+h_{0}\ ,
\end{equation*}
with which the equation (\ref{mxeqcrystal}) becomes 
\begin{equation}
b_{1}^{\prime \prime }+\frac{K}{8}(\omega -4b_{1})\sin ^{2}(H)\Big(%
l_{1}^{2}\left( \lambda \omega ^{2}-16\right) -\frac{l_{1}^{2}}{l_{3}^{2}}%
\lambda +\frac{\lambda l_{1}^{2}}{l_{3}^{2}}(\omega ^{2}l_{3}^{2}-1)\cos (2H)%
\Big)=0\ .
\end{equation}

In terms of the same variables of Eq. (\ref{newcoord}), Eq. (\ref%
{mxeqcrystal}) can be written as a form of confluent Heun's equation as in
the previous section. That is, 
\begin{equation}
\frac{d^{2}y}{dx^{2}}+\Big(8K\lambda \ \sin ^{2}x-\Delta ^{2}\sin ^{2}2x\Big)%
y=0\ ,  \label{eqytc}
\end{equation}%
with a non-negative constant $\Delta \geq 0$, 
\begin{equation*}
\Delta ^{2}:=\frac{K\lambda ^{2}}{4}\Big(\omega ^{2}-\frac{1}{l_{3}^{2}}\Big)%
\ .
\end{equation*}%
\newline
In this section, we assume that $\omega ^{2}\geq 1/l_{3}^{2}$. The equation (%
\ref{eqytc}) can be cast into the confluent Heun's equation 
\begin{equation}
\frac{d^{2}}{dz^{2}}Y(z)+\Big(\frac{\gamma }{z}+\frac{\delta }{z-1}+\epsilon %
\Big)\frac{d}{dz}Y(z)+\frac{\alpha z-q}{z(z-1)}Y(z)=0\ ,
\end{equation}%
where 
\begin{eqnarray}
&&z=\cos ^{2}x\ ,\qquad Y(z)=e^{-\Delta z}\ y(\arccos \sqrt{z})\ ,  \notag \\
&&\gamma =\delta =1/2\ ,\qquad \epsilon =2\Delta \ ,\qquad \alpha =\Delta
+2K\lambda \ ,\qquad q=\Delta /2+2K\lambda \ .
\end{eqnarray}%
A general solution to this equation is known as 
\begin{eqnarray}
&&Y(z)=C_{1}\ \text{HeunC}(\Delta+2K\lambda, 1/2, 1/2, 2\Delta, \Delta
/2+2K\lambda ;z)  \notag \\
&&\hspace{0.5in}+C_{2}\sqrt{z}\ \text{HeunC}(2\Delta+2K\lambda, 3/2, 1/2,
2\Delta, 3\Delta /2+2K\lambda -1/4 ;z)\ .
\end{eqnarray}%
The confluent Heun's function can be expanded in terms of Kummer's confluent
functions when $\epsilon \neq 0$, and $\gamma +\delta $ is not zero nor
negative integer \cite{ishkhanyan}. Our equation satisfies this condition so
that 
\begin{equation}
Y(z)=\sum\limits_{n=1}^{\infty }a_{n}\ {}_{1}F_{1}(n+\frac{1}{2};\ \frac{1}{2%
};\ -2\Delta z)\ ,
\end{equation}%
where ${}_{1}F_{1}$ is the Kummer's confluent hypergeometric function, and
the coefficients are determined by the recursion relation 
\begin{eqnarray}
&&n\Big(n-\frac{K\lambda }{\Delta }\Big)a_{n}+\Big\{-2n^{2}+\Big(\frac{%
2K\lambda }{\Delta }+2\Delta +3\Big)n-\Big(\frac{3K\lambda }{2\Delta }%
+2K\lambda +\frac{3\Delta }{2}+\frac{5}{4}\Big)\Big\}a_{n-1}  \notag \\
&&+\Big(n-\frac{3}{2}\Big)\Big(n-\frac{3}{2}-\frac{K\lambda }{\Delta }\Big)%
a_{n-2}=0\ .
\end{eqnarray}%
It is worth to notice that this kind of series is terminated if 
\begin{eqnarray}
&&\frac{K\lambda }{\Delta }=N+\frac{1}{2}\ ,  \label{quantcond1} \\
&\Rightarrow &\ \omega (N)=\pm \Big(\frac{1}{l_{3}^{2}}+\frac{4K}{(N+1/2)^{2}%
}\Big)^{1/2}\ ,  \label{quantcond2}
\end{eqnarray}%
for some natural number $N\in \mathbb{N}$.

A possible criticism to the time-crystals constructed in the previous
references \cite{Fab1}, \cite{gaugsk} is that there was no argument to fix
the corresponding time-periods. It is a very intriguing results that the
classic theory of Kummer's confluent functions is able to fix the
time-period of the present gauged time-crystals based on the Heun equation
through the quantization condition in Eq. (\ref{quantcond2}).

\subsubsection{Relation with the Whittaker-Hill equation}

Following the same steps as the gauged Skyrmion, the mapping with the
Whittaker-Hill equation 
\begin{equation*}
y^{\prime \prime }+\left( \lambda _{0}+4\alpha s\cos {(2x)}+2\alpha ^{2}\cos 
{(4x)}\right)y =0\ ,
\end{equation*}%
determines the coefficients as 
\begin{equation*}
\lambda _{0}=\frac{K\lambda }{8l_{3}^{2}}\left( \lambda -l_{3}^{2}(\lambda
\omega ^{2}-32)\right) \ ,\quad \alpha ^{2}=\frac{K\lambda ^{2}}{16l_{3}^{2}}%
\left( l_{3}^{2}\omega ^{2}-1\right) \ ,\quad s=-4l_{3}\sqrt{\frac{K}{%
l_{3}^{2}\omega ^{2}-1}}\ ,
\end{equation*}%
so that, the resurgence parameter is given by 
\begin{equation*}
g^{2}=l_{3}\left( \lambda K(l_{3}^{2}\lambda \omega ^{2}-8l_{3}^{2}-\lambda
)\right) ^{-1/2}\ .
\end{equation*}

\section{Conclusions and perspectives}

\label{conclusions}

We have shown that one can get a complete analytic description of gauged
Skyrmions in (3+1) dimensions living within a finite volume in terms of
classic results in the theory of ordinary differential equations. In
particular, we have been able to reduce the coupled field equations of the
gauged Skyrme model (which, in principle, are seven coupled non-linear PDEs)
in two non-trivial topological sectors (one corresponding to gauged
Skyrmions and the other to gauged time-crystals) to the Heun equation
(which, for some particular choice of the parameters, can be further reduced
to the Whittaker-Hill equation). This technical result has many intriguing
consequences. First of all, one obtains a complete explicit construction of
these gauged solitons in terms of Heun and Kummer functions (so that, for
instance, it is possible to compute the energy of the system in terms of the
Baryon charge and the volume of the region). Secondly, the time-period of
the time-cystals is quantized. Likewise, the volume occupied by the gauged
Skyrmions is quantized. The present analysis also discloses the appearance
of resurgent phenomena within the gauged Skyrme model in (3+1) dimensions.
In particular, suitable electromagnetic perturbations of the gauged
Skyrmions satisfy the Mathieu equation (which is a well known example in
which the resurgent paradigm works very well). Thus, the spectrum of these
perturbations can be determined in terms of known results in the theory of
the Mathieu equation.

It is worth to further analyze the appearance of resurgent phenomena in the
Skyrme model as this analysis could help to shed new light on resurgence in
QCD as well. We hope to come back on this important issue in a future
publication.

\subsection*{Acknowledgements}

M.L. and A.V. appreciates the support of CONICYT Fellowship 21141229 and
21151067, respectively. This work has been funded by the FONDECYT grants
1160137 and 1181047. The Centro de Estudios Cient\'{\i}ficos (CECs) is
funded by the Chilean Government through the Centers of Excellence Base
Financing Program of CONICYT. The work is supported in part by National
Research Foundation of Korea funded by the Ministry of Education (Grant
2018-R1D1A1B0-7048945).

\end{document}